\documentclass[oneside, 11pt]{article}

\usepackage[dvips]{graphicx}
\usepackage{amsmath, comment}
\usepackage{amssymb}
\usepackage{sectsty}
\usepackage{titlesec}
\usepackage[margin=1in]{geometry}
\usepackage{fancyhdr}
\usepackage{color, colortbl}
\usepackage{amsthm}
\usepackage{xcolor,colortbl}
\usepackage{enumitem}
\usepackage{bm}
\usepackage{ulem}
\usepackage{graphicx}
\usepackage{subfig}

\definecolor{Gray}{gray}{0.85}
\definecolor{LightCyan}{rgb}{0.88,1,1}
\newcolumntype{a}{>{\columncolor{Gray}}c}

\theoremstyle{plain}

\newtheorem*{thm*}{Theorem}

\allsectionsfont{\bf \large }

\titlespacing*{\section}{0pt}{12pt}{3pt}{}
\titlespacing*{\subsection}{0pt}{12pt}{3pt}{}

\newcommand{\ben}{\begin{enumerate}}
\newcommand{\een}{\end{enumerate}}
\newcommand{\bit}{\begin{itemize}}
\newcommand{\eit}{\end{itemize}}

\definecolor{Gray}{gray}{0.9}
\newcolumntype{g}{>{\columncolor{Gray}}c}
\allowdisplaybreaks

\usepackage[square,numbers]{natbib}
\def \ts {{\,}}

\def \hbar {{\bar h}}

\titleformat{\section}{\normalfont\normalsize \bfseries }{\thesection.}{0.5em}{}
\titlespacing*{\section}{0em}{1.0em}{-0.75em}

\begin{document}

\begin{center}{\Large Future Evolution of COVID-19 Pandemic in North Carolina:
\break
Can We Flatten the Curve?}
\\
\vspace{4mm}
\normalsize
Omar El Housni$^1$\footnote{Email addresses of the authors are 
oe2148@columbia.edu, 
ms3268@cornell.edu, 
rusmevic@marshall.usc.edu, 
topaloglu@orie.cornell.edu, and  
ziya@email.unc.edu.}\!\!, Mika Sumida$^2$, Paat Rusmevichientong$^3$, Huseyin Topaloglu$^2$, Serhan Ziya$^4$
\\
\vspace{2mm}
\footnotesize
$^1$Department of Industrial Engineering and Operations Research, Columbia University, New York, NY 10027
\\
$^2$School of Operations Research and Information Engineering, Cornell Tech, New York, NY 10044
\\
$^3$Marshall School of Business, University of Southern California, Los Angeles, CA 90089
\\
$^4$Department of Statistics and Operations Research, University of North Carolina, Chapel Hill, NC 27599
\\
\normalsize
\vspace{1mm}
\today
\end{center}

\vspace{-7mm}

\begin{abstract}
\noindent On June 24th, Governor Cooper announced that North Carolina will not be moving into Phase 3 of its reopening process at least until July 17th. Given the recent increases in daily positive cases and hospitalizations, this decision was not surprising. However, given the political and economic pressures which are forcing the state to reopen, it is not clear what actions will help North Carolina to avoid the worst. We use a compartmentalized model to study the effects of social distancing measures and testing capacity combined with contact tracing on the evolution of the pandemic in North Carolina until the end of the year. We find that going back to restrictions that were in place during Phase 1 will slow down the spread but if the state wants to continue to reopen or at least remain in Phase 2 or Phase 3 it needs to significantly expand its testing and contact tracing capacity. Even under our best-case scenario of high contact tracing effectiveness, the number of contact tracers the state currently employs is inadequate. 
\end{abstract}

\section{Background}
\label{sec:mainbody}

This report provides a summary of our analysis into how social distancing measures along with testing and contact tracing capacity will likely impact the evolution of COVID-19 pandemic within the state of North Carolina until the end of 2020. This analysis is based on a model, which was first introduced in~\cite{ElHousni2020}, a report on the findings of a similar study conducted for New York City. For the convenience of the reader, this paper is prepared as a self-contained document that includes a complete description of the mathematical model and therefore parts of it which pertain to the mathematical model in general overlap with~\cite{ElHousni2020}. 

The first COVID-19 case in North Carolina was identified on March 3rd and the state of emergency was declared on March 10th. On March 30th, statewide stay-at-home order was put in place and non-essential businesses were ordered closed. These orders stayed in place until May 8th. Starting with that date, the state entered Phase 1 of its reopening process. On May 22nd, Phase 2 took effect with a scheduled end date of June 26th. However, on June 24th, Governor Cooper announced that this date has been pushed three weeks into the future, to July 17th. It is not clear whether the state will indeed move to Phase 3 on that date or it will be postponed again.

Over the last few weeks, North Carolina has experienced significant increases in the number of new cases and hospitalizations as the state has continued to lift restrictions. At the same time, pressures that have forced the officials and other administrators to open up the economy (economical, political, or otherwise) appear to be still there and it is not clear what course of action will help balance the competing priorities. Our objective is to investigate the impact of different social distancing measures the state might choose to put in place in the future, whether increases in state's testing capacity might help in mitigating the impact of reopening on hospitalizations and deaths, and what resource demands will be needed for contact tracing. In the following, we summarize our main findings. For a broad description of our mathematical model, see Section \ref{sec:model}. For details, see the Appendix.

\section{Move Forward to Phase 3, Stay in Phase 2, or Go Back to Phase 1?}

\begin{figure}[!htbp]
\begin{center}
30,000 tests per day
\\
\includegraphics[scale=0.4]{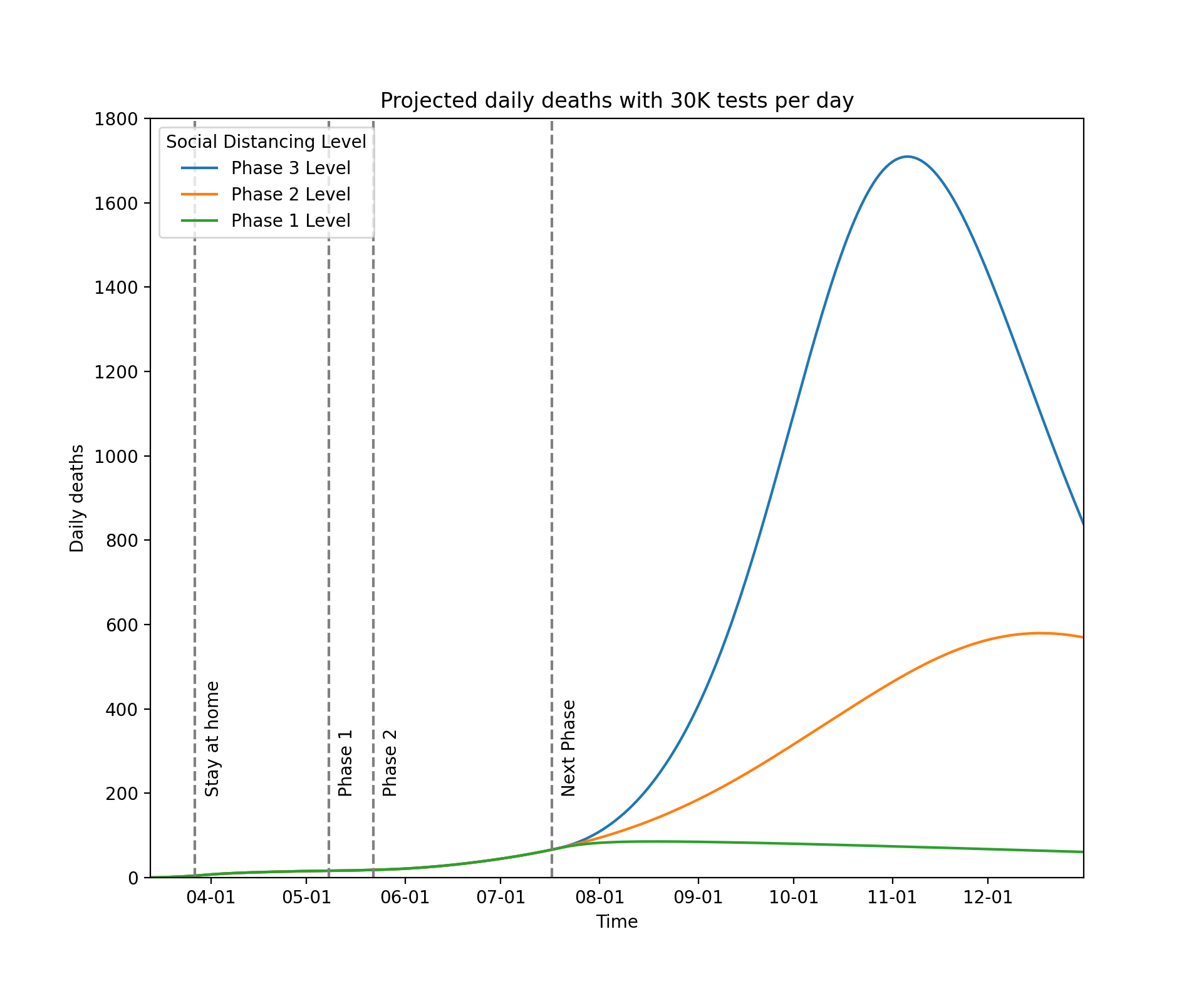}
\vspace{2mm}
\\
60,000 tests per day
\\
\includegraphics[scale=0.4]{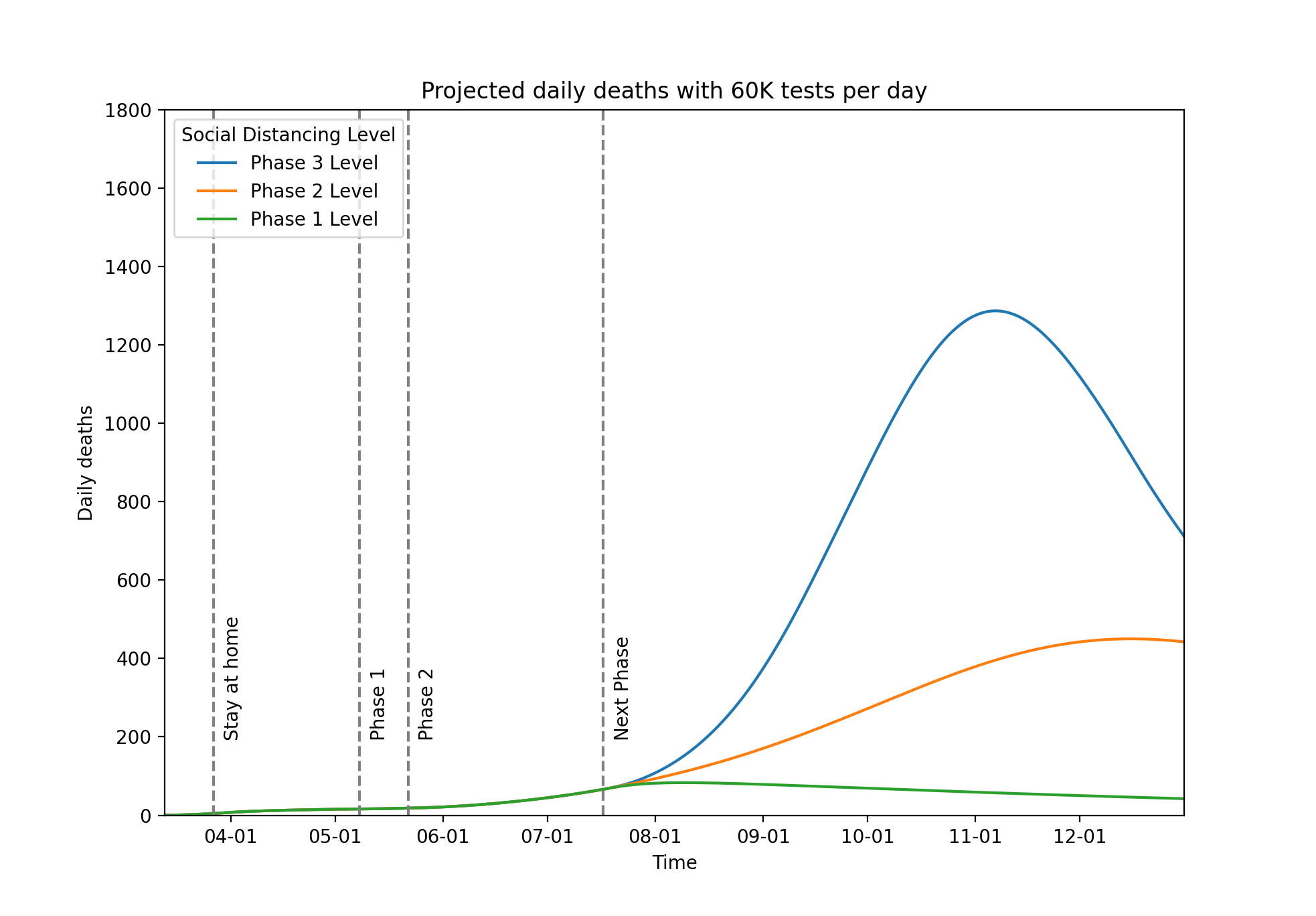}

\caption{Trajectory of the daily deaths under testing capacities of 30,000 and 60,000  per day.}
\vspace{-8mm}
\label{fig:traj}
\end{center}
\end{figure}

We do not know the exact current testing capacity in North Carolina, however, the data indicate that the maximum number of tests performed on a single day is between 25,000 and 30,000. Therefore, we chose a testing capacity of 30,000 per day as the base case. We also do not know what the testing capacity will be in the future but to get a sense of the impact of the test availability on controlling the spread of the disease, we considered 60,000 per day as an alternative scenario. The top and bottom panels of Figure \ref{fig:traj} respectively provide the trajectory of projected daily deaths under the testing capacity of 30,000 per day and 60,000 per day. Each data series in the two panels corresponds to a different level of social distancing that will be practiced starting with July 17th. We consider three possibilities: moving to Phase 3, staying in Phase 2, or going back to Phase 1 all assumed to be in place until the end of the year. It is important to note that we do not expect the state to stick with any one of these three choices until the end of the year, however, comparing the projection under each scenario provides some insights into what kind of impact each one of these choices will have on the disease's spread. 

As we can see from the blue curve in the top panel of the figure, with a testing capacity of 30,000 tests per day, moving to Phase 3 will likely be disastrous with a major peak that will arrive some time in early November and a projected number of deaths of approximately 160,000 by the end of the year. (Labels on the right end of each panel give the total number of deaths by the end of 2020.) In fact, such a major peak would mean that hospitals would be overwhelmed to such a level that they may not be able to provide their regular level of care and the mortality rates might end up being more than what the data to date suggest and our analysis assumes. If the state chooses to stay in Phase 2, there will surely be fewer deaths, however, we predict that there will still be a significant increase in daily deaths and hospitalizations compared with current levels with a peak in December. 

From the bottom panel, we can see that having more tests would definitely help resulting in fewer deaths but even a capacity of 60,000 tests per day will not be sufficient to prevent a major wave of hospitalizations and deaths in Fall. The figure suggests that what will help in reining in the spread of the disease is going back to the restrictions that were in place in Phase 1 regardless of whether the testing capacity is 30,000 or 60,000. It appears that increasing testing capacity (at least to 60,000) will not help mitigate the impact of relaxing social distancing measures.

\section{A Closer Look into Testing Capacity and Social Distancing: What Might Help?}

The discussion in the previous section suggests that even if the state does not move into Phase 3 and stays in Phase 2 until the end of the year, hospitalizations and deaths will soar over the next few months even if the daily testing capacity increases to 60,000. Moving back to Phase 1 would help but given the highly likely economic impact of taking such an action it might be helpful to consider alternatives. In particular, it is of interest to investigate what levels of testing capacity might be sufficient for different levels of social distancing that might possibly be experienced in the coming months as a result of different policies that might be employed. 

\begin{figure}[t]

\begin{center}
\includegraphics[scale=0.55]{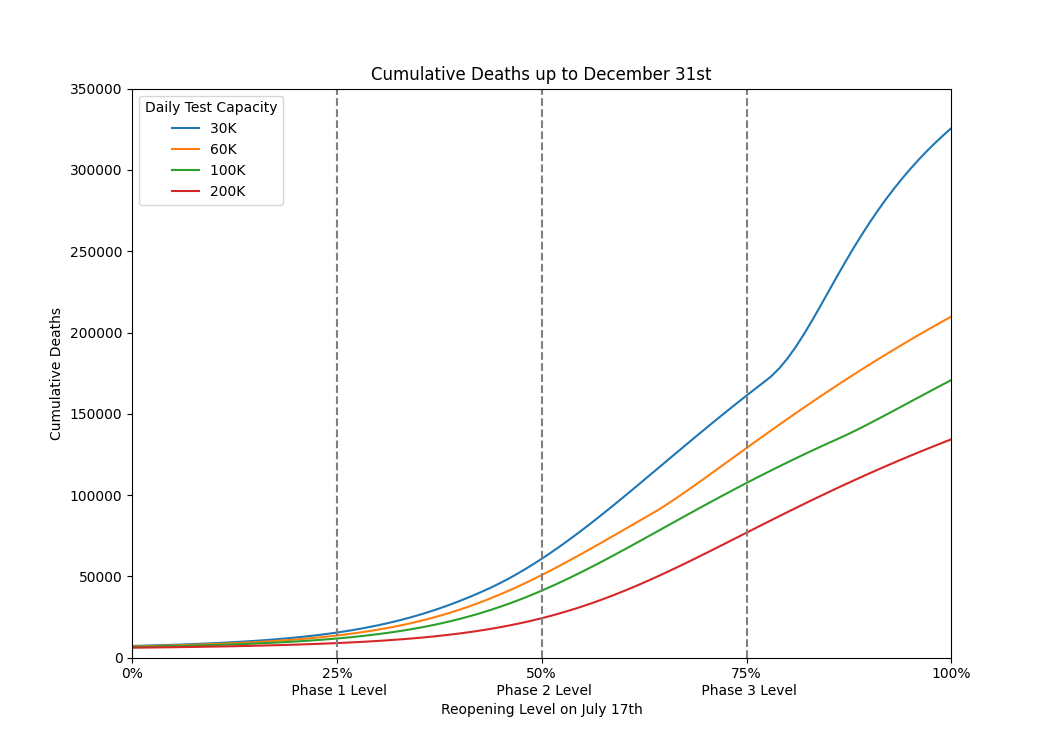}
\vspace{2mm}
\\
~\hspace{-2.5mm}
\caption{Trade-offs between the deaths by the end of 2020 and the social distancing norms.}
\vspace{-4mm}
\label{fig:front}
\end{center}
\end{figure}

Figure \ref{fig:front} shows the trade-offs between the total number of deaths by the end of the year and the level of social distancing to be practiced post July 17th, under different testing capacities. The horizontal axis shows the level of social distancing, expressed as a percentage relaxation in the social distancing norms in reference to what was experienced before Phase 1 under the stay-at-home order. Thus, 0\% corresponds to stay-at-home, whereas 100\% is full relaxation, corresponding to life before the pandemic with no restrictions in place. Social distancing levels that correspond to Phases 1, 2, and 3 are also marked on the x-axis so that the reader can get a sense of roughly what different social distancing levels correspond to. (Social distancing levels in Phase 1, 2, and 3 respectively correspond to 25\%, 50\%, and 75\%, which is in line with social distancing estimates in~\cite{ihme2020}.) The vertical axis shows the total number of projected deaths by the end of 2020. Each data series corresponds to a different level of testing capacity. Our analysis assumes that new testing capacity and social distancing norms become effective on July 17th. In computing these projections, we use the reported number of tests performed on each day up to June 15th. Between June 15 and July 17 we assume 30,000 tests are conducted daily.

Current discussions and political and economical climate in the state suggest that going back to Phase 1 is not being considered as a serious possibility at least in the near future. In fact, with planned opening of the schools and colleges in late summer, the state is on path to further opening. This suggests that, it would make sense to pay more attention to the right half of Figure \ref{fig:front}, the part that corresponds to relaxation levels that are larger than 50\%, i.e., either staying at Phase 2 or moving further to Phase 3 and beyond. There, we can observe that having more tests available everyday would make a significant difference. To be clear, having more tests cannot be a substitute for social distancing. As we can see from the figure, even under the daily testing capacity of 200,000, the total number of deaths by the end of the year would be very large. Social distancing appears to be the only way to keep the spread under control. Still, it is important to note that even if more tests will likely not alter the dynamics of the disease's spread it will at least help in keeping the numbers smaller.


\section{Testing Capacity and the Effectiveness of Contact Tracing}

One important point to highlight about our model is that it assumes that there is a contact tracing policy in place. More specifically, it assumes that close contacts of individuals who test positive are identified and they are asked to self-isolate, which helps in reducing new infections in the population. In fact, this is the main reason why in our model more testing helps. Without contact tracing, the benefit of having more tests beyond a certain level would be minimal. Our analysis makes certain assumptions about the effectiveness of contact tracing but there is in fact significant uncertainty around this issue. To partially address this uncertainty, we consider alternative scenarios in regards to contact tracing effectiveness and investigate the changes in our projections. (As we explain in the Appendix, because we are not aware of any studies on contact tracing in the United States, we used a study conducted in China to come up with a rough approximation for a parameter that captures the contact tracing effectiveness.)

We control the effectiveness of contact tracing via the parameter CTEP, which stands for \textit{Contact Tracing Effectiveness Parameter}. For details on CTEP, we refer the reader to the Appendix, where $\tt likOfBeingInfected$ is used for CTEP, but it might be useful for the reader to know that CTEP is basically an indicator of how successful contact tracing is in identifying asymptomatic infected individuals over the non-infected ones. If CTEP = 1, this means that contact tracing is of no use at all and higher values of CTEP indicate higher likelihood of identifying infected people or more effective contact tracing.

\begin{figure}[!htbp]

\begin{center}
\includegraphics[scale=0.4]{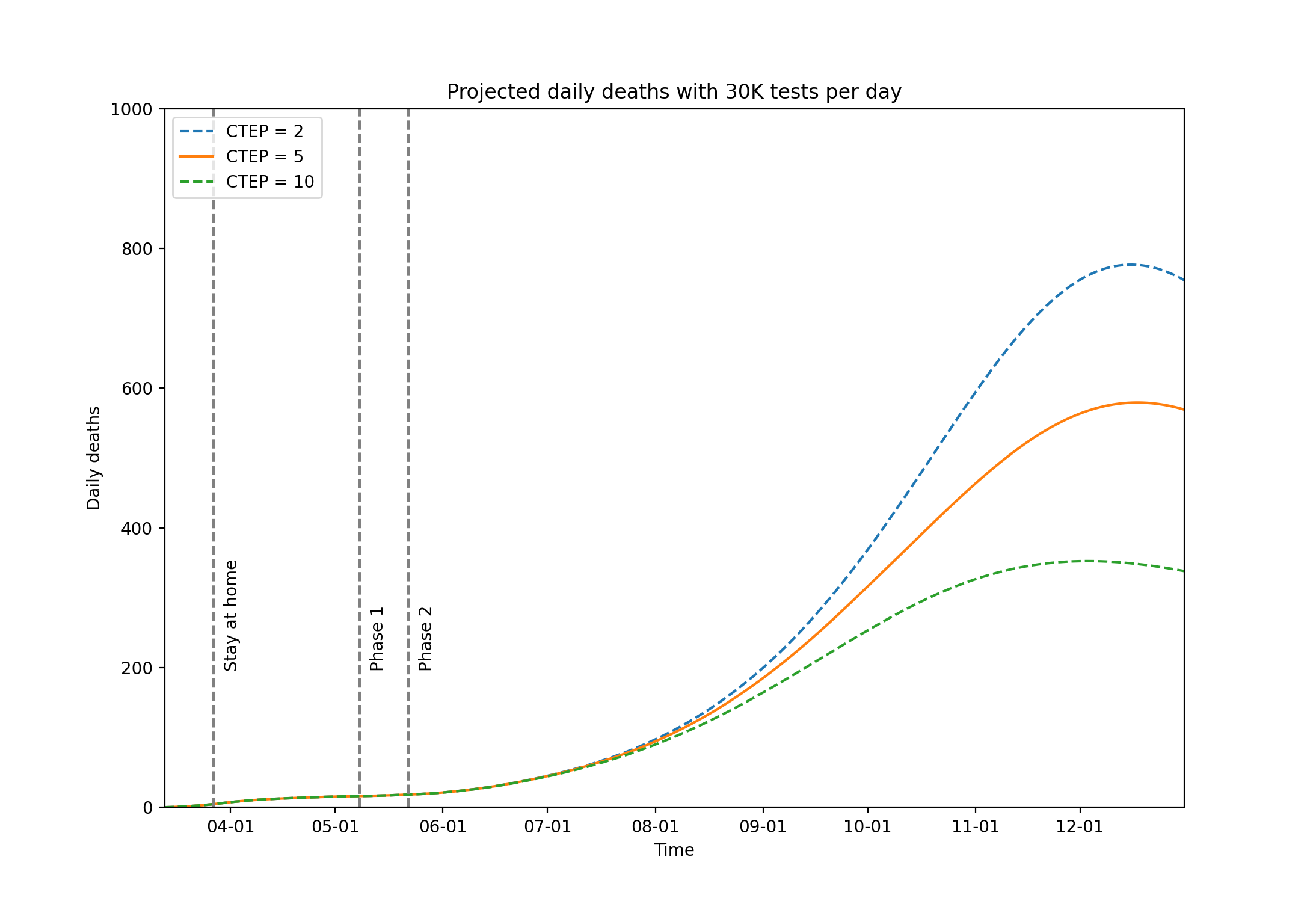}
\vspace{2mm}

\caption{Impact of Contact Tracing Effectiveness on Projections for the Daily Number of Deaths.}
\vspace{-4mm}
\label{fig:CTEP}
\end{center}
\end{figure}

Our analysis above assumed CTEP = 5 based on our rough approximation for the parameter. Figure \ref{fig:CTEP} considers two additional cases, one with CTEP = 2 and the other with CTEP = 10 all under the assumption that the state will continue to operate in Phase 2 until the end of the year. 

Not surprisingly, when contact tracing is more effective, number of deaths decline considerably. Nevertheless, even when CTEP = 10, we project a very large number of deaths and hospitalizations by the end of the year with a major peak in late Fall. However, it is important to remind the reader that, in fact, we do not know how effective contact tracing is or will be in the future and it is possible that even setting CTEP to 10 might be underestimating the actual effectiveness. We do hope that is the case as that would suggest that, given the substantial decreases in the projected number of deaths with increased contact tracing efficiency and perhaps combined with increased testing capacity, the state might be able to keep the spread somewhat under control. The critical question in that case would be whether the state has the sufficient resources for contact tracing. This is what we investigate next.

\section{Projected Resource Needs for Contact Tracing}

\begin{figure}[!htbp]
\begin{center}
\includegraphics[scale=0.4]{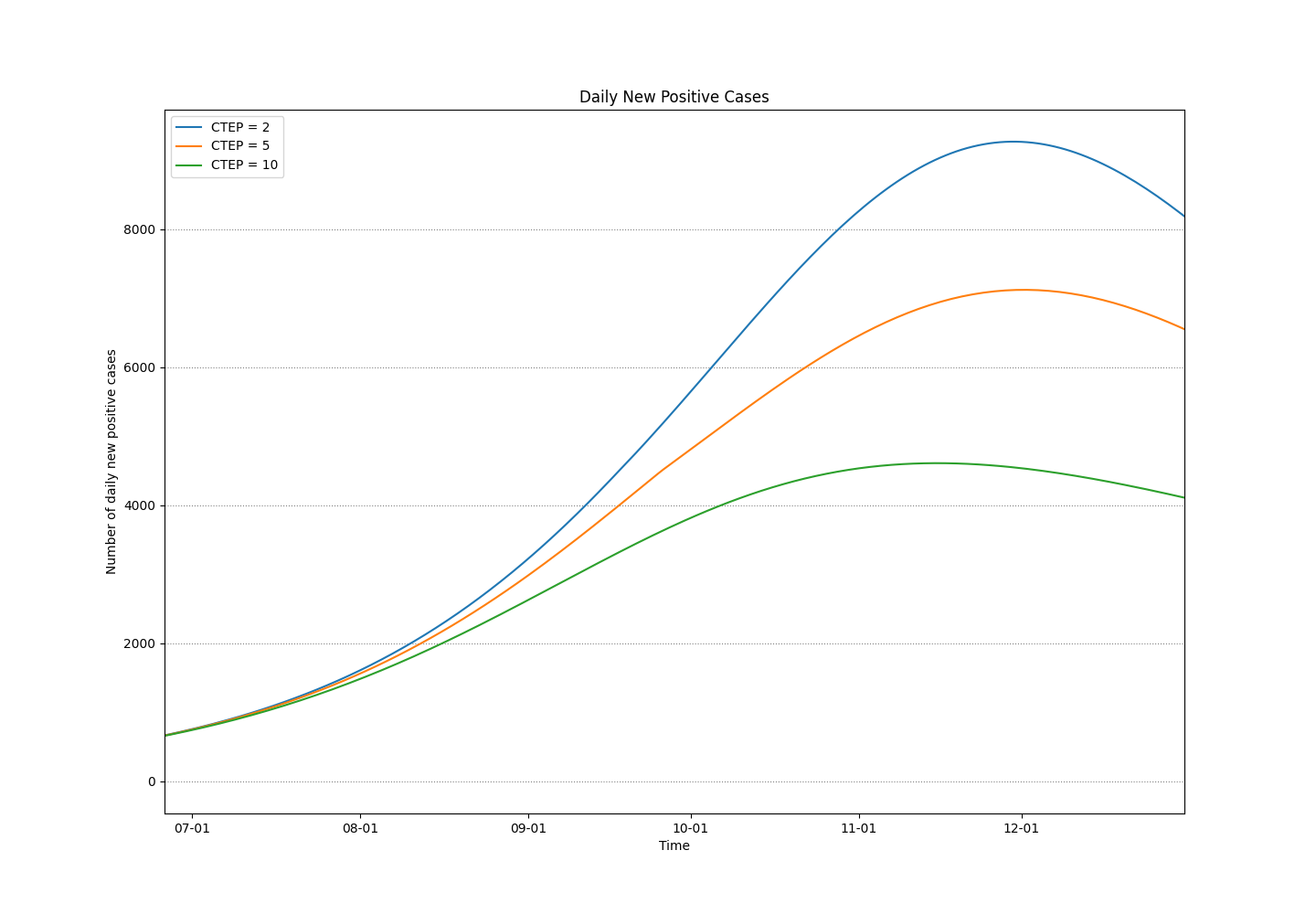}
\vspace{2mm}
\caption{Projections of Daily Number of New Positive Cases.}
\vspace{-4mm}
\label{fig:CT1}
\end{center}
\end{figure}

\begin{figure}[!htbp]
\begin{center}
\includegraphics[scale=0.4]{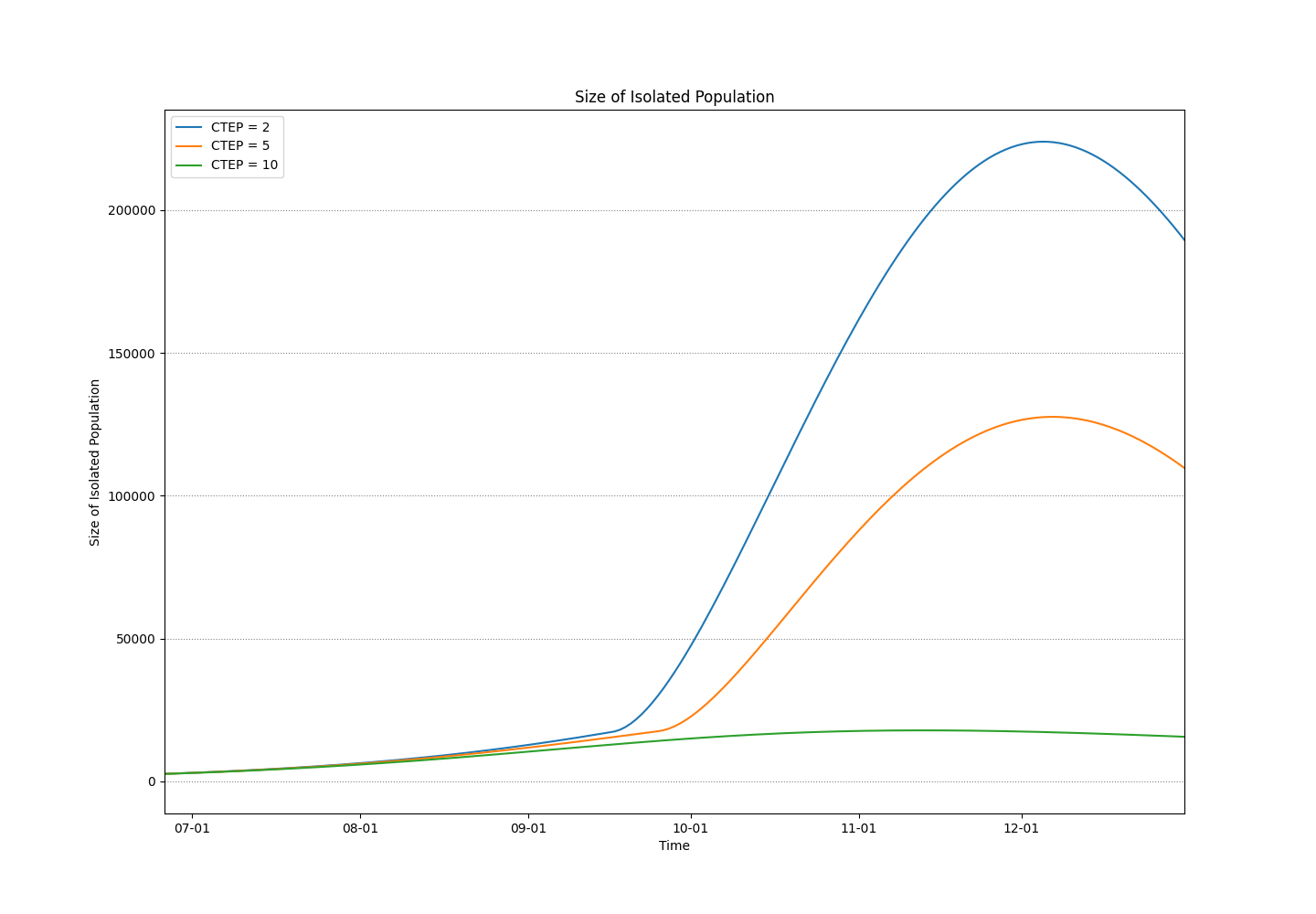}
\vspace{2mm}
\caption{Projections of Daily Number of Isolated Individuals who Need Follow-up.}
\vspace{-4mm}
\label{fig:CT2}
\end{center}
\end{figure}

There are mainly two components of contact tracing. First, close contacts of individuals who test positive are identified and tracers contact these individuals informing them about the possibility of their having been infected and giving them directions as to what they need to do over a period of roughly two weeks. Second, during this period, when these close contacts need to self-isolate (unless they are tested and result comes negative), tracers periodically contact them to check up on their conditions and ensure that they are self-isolating. Figure \ref{fig:CT1} shows our daily projections for new cases, who need to be interviewed to determine their close contacts while Figure \ref{fig:CT2} shows our projections for the number of individuals who need follow-up calls from contact tracers.

Again, we can see from the figures that contact tracing resource needs peak some time in Fall regardless of how effective tracing is, however, the effectiveness has a significant impact on the number of tracers that would be needed. For the mid-range effectiveness scenario, where CTEP = 5, the maximum value for the number of daily new positive cases is 7,121 whereas the maximum value for the number of isolated individuals who need follow-up is 127,617. According to the estimates by the Fitzhugh Mullan Institute for Health Workforce Equity of the George Washington University, a single contact tracer, assuming an 8-hr workday, would be able to interview six index cases (individuals who test positive), or make 12 initial contact notifications, or make 32 follow-up calls~\cite{GWHWI2020}. This would then mean that North Carolina would need roughly 7,550 contact tracers. (This number is very close to 7,700, the current estimate by the Fitzhugh Mullan Institute). Our estimate would be roughly 11,630 for CTEP = 2, and 2,860 for CTEP = 10. According to recent reports~\cite{NO-Contact}, North Carolina has roughly 1,500 contact tracers meaning that even under our more optimistic scenario of high contact tracing effectiveness, the state's tracing resources would fall short of meeting the demand.

\section{Discussion}

 Our main objective in this work has not been to make precise predictions for key metrics like number of deaths, number of hospitalizations etc.\ but rather to understand the dynamics of the spread of the disease, how these dynamics would depend on the state's testing and contact tracing capacity, and provide insights into whether the existing capacity is sufficient for the state to reopen without overwhelming its healthcare resources. Therefore, the reader should take the precise values of the estimates we report in this paper with a grain of salt but give credence and pay more attention to the insights we generated through their analysis collectively. Unfortunately, this analysis strongly suggests that even if North Carolina does not move to Phase 3 on July 17th, without further restrictions, the number of new cases, hospitalizations and deaths will continue to increase substantially peaking some time in Fall.

 There are several ways that this grim prediction may not become reality. One possibility is for the state to reimpose some of the restrictions that were in place in Phase 1 of the reopening. Alternatively, of course, the state might continue with is reopening plan but take other actions that would significantly limit person-to-person transmission such as enforcing face covering requirements more strictly. The goal here should be to bring down the transmission rate at least to the levels achieved under Phase 1. The current economic and political climate in North Carolina suggests that reimposing restrictions is not very likely and it is not clear whether strict requirements for face coverings could keep new infections down as the state keeps opening.
 
 If the state continues with its reopening plan with no further restrictions, there appears to be only one way the spread of the disease can be kept under control and that is through extensive testing combined with effective contact tracing. There is some uncertainty as to how effective contact tracing will be but it is safe to say that North Carolina is significantly understaffed to keep up with the workload that will be generated for contact tracers over the next few months. Therefore, it is essential for the state to quickly hire new contact tracers to keep up with the impending jump in their workload and ensure that contact tracing is effective in identifying infected individuals and prevent them from infecting others in the community. It is also important to understand that testing capacity and contact tracing capacity are directly tied to each other and any meaningful response should include increasing both in some proportional manner. Beyond certain levels of testing capacity, further increases will not bring much benefit if contact tracers are already overwhelmed. Similarly, hiring more contact tracers will be meaningless unless the state continues to test at certain numbers everyday. 
 
Without further action from the state, one possibility that could alter our predictions is that, with the increased public awareness around how the disease spreads, the infection rates for different phases might end up being lower than what we assumed in our analysis based on the mobility data. It is also possible that over the next few months, the disease's spread might follow a course that is quite different from what we observed so far. This may not necessarily be a result of some form of mutation in the virus but could simply be a result of the changes in the demographics of the population who is infected by the virus over time. In fact, our analysis of the most recent data from North Carolina suggests that there appears to be some drop in the overall mortality rate and this appears to be a national trend. Recent news articles reported that median age of individuals who test positive has dropped substantially since March~\cite{nyt0625} and such a change in the infected population would reduce hospitalization and mortality rates. (Interestingly, in North Carolina, even though we observe a drop in mortality rates, hospitalization rate appears to not have changed in any significant manner.) If this trend continues, it is possible that the scale of the problem we will face will be smaller than what we expect based on our analysis. However, it is important to keep in mind that younger individuals who tend to be mostly asymptomatic and are more socially active, can spread the virus more easily to different segments of the population, which might increase the hospitalization and mortality rates back up again. In any case, it would be prudent to not put our hopes on possibilities that are beyond our control but take actions that would help even under the worst circumstances.

\section{Outline of the Mathematical Model}
\label{sec:model}
Our model follows the standard susceptible-infected-recovered paradigm with compartments capturing individuals classified along the dimensions of infected, noninfected, and recovered, as well as symptomatic-isolated, asymptomatic-isolated, and asymptomatic-nonisolated. Once an individual is tested positive, a certain number of individuals who are expected to have been in close contact with the positive individual are transferred to a certain isolated compartment. In this way, we build a contact tracing mechanism to ensure that those who have been in contact with a positive individual reduce their contact with the rest of the population. Running our model with a starting date of March 2, we get a trajectory of the pandemic that is in reasonably close agreement with the actual trajectory in North Carolina so far. In the appendix, we give a discussion of our model and provide comparisons of its output with the actual trajectory of the pandemic.

\bibliographystyle{abbrvnat}
\bibliography{references}

\newpage
\appendix

\appendix
\noindent {\bf \Large Appendix}
\\
The appendix is organized as follows. In Appendix \ref{section:model}, we discuss our model at a high level.~This section avoids any technical details but it is useful for seeing how our model works, what compartments it uses, and how the different compartments interact with each other. In Appendix~\ref{section:system-dynamics}, we provide the full mathematical details of our model. In Appendix \ref{section:model-parameters}, we elaborate on the parameters of our model and how they are estimated. We borrow some of the parameters from the literature and derive others based off of problem primitives. We explain how we arrive at the derived model parameters. In Appendix \ref{section:model-validation}, we compare the output of our model with the trajectory of the pandemic in North Carolina until early May and demonstrate that our model does a reasonably good job of predicting the trajectory that has been observed so far.

\section{Model at a High Level}
\label{section:model}

\vspace{10pt}

In our model, we capture the two benefits of testing, which are isolating the specific individual tested positive and tracing the contacts of the positive individual. In particular, our model distinguishes the infected individuals who know that they are infected from the infected individuals who do not know whether they are infected. Once a person is tested positive, this person is an infected individual who knows that she is infected. Such a person minimally infects others. Moreover, considering the individuals who do not know whether they are infected, we classify them as symptomatic-isolated, \mbox{asymptomatic-isolated}, or \mbox{asymptomatic-nonisolated}. Once an individual is tested positive, a certain number of people around this person are classified as~\mbox{asymptomatic-isolated} people, so these individuals  reduce their contact with the general population, preventing them from infecting others to some extent.

\begin{figure}[t]
\begin{center}
~
\includegraphics[scale=0.46,trim=2cm 5cm 0cm 1cm]{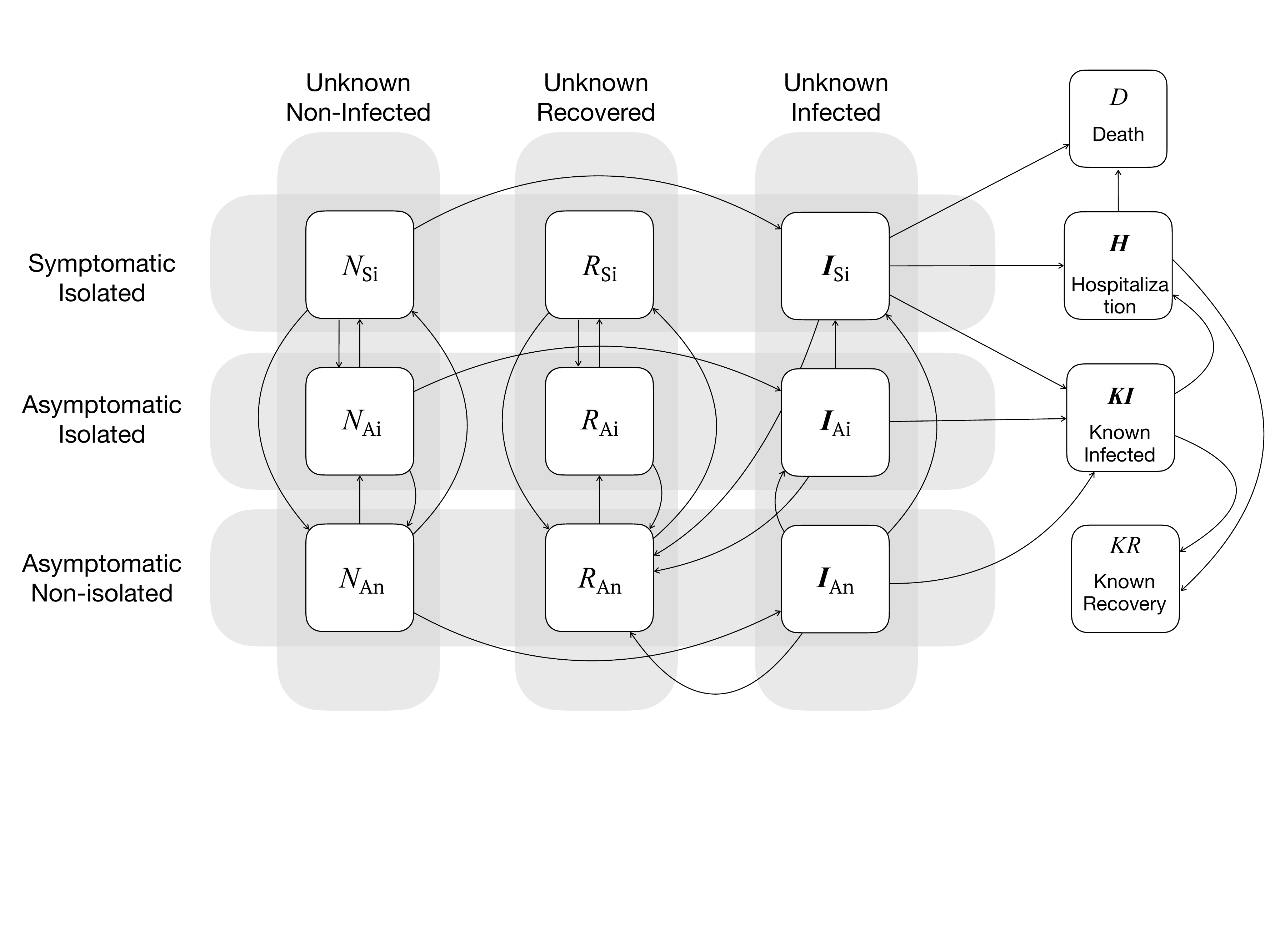}
\vspace{-5mm}
\caption{Compartments used in our model.}

\label{fig:comp}
\end{center}
\end{figure}

Our model follows the standard susceptible-infected-recovered modeling paradigm. In Figure~\ref{fig:comp}, we show the compartments in our model. The boxes represent the compartments. The arrows represent the possible flows between the compartments. The three compartments on the right of the figure capture the  individuals who know that they are infected. In particular, the compartment labeled ``KI'' corresponds to the infected individuals who know that they are infected and not (yet) hospitalized. The compartment labeled ``H'' captures the infected individuals who are hospitalized. The individuals in the latter compartment also know that they are infected. The compartment labeled ``KR'' captures the recovered individuals who knew that they were infected.~Perhaps optimistically, in our model, the individuals in all of these three compartments minimally infect others, since they are aware that they were infected.

The nine compartments on the left side of the figure capture the individuals who do not know whether they are infected. Some of these individuals will, in actuality, be infected, some will be noninfected, and some will even have recovered without knowing they had been infected.~We classify the individuals who do not know whether they are infected along two dimensions. In the first dimension, any such individual could be symptomatic-isolated, asymptomatic-isolated, or asymptomatic-nonisolated. Asymptomatic-nonisolated individuals do not show symptoms of the disease and do not make an effort to reduce their contact. Asymptomatic-isolated individuals make a conscious effort to reduce their contact, mainly because they have been in touch with a person who tested positive.~Symptomatic-isolated individuals show symptoms but do not know whether they are infected. Their symptoms may be due to other diseases. 

Once again, perhaps optimistically, in our model, all symptomatic individuals isolate themselves, so we do not have a compartment for symptomatic-nonisolated people. Symptomatic-isolated and asymptomatic-isolated individuals infect others with a smaller rate, when compared with asymptomatic-nonisolated people. When stricter social distancing norms are enforced, all individuals will reduce their contact with the general population, but we follow the assumption that asymptomatic-nonisolated individuals will still maintain a higher contact rate with the rest of the population than asymptomatic-isolated and symptomatic-isolated individuals.

Considering the individuals who do not know whether they are infected, we classify them along another dimension. In this second dimension, an individual who does not know whether she is infected could be infected, noninfected, or recovered. Thus, considering the nine compartments on the left side of Figure \ref{fig:comp}, the compartment labeled ``$I_{\text{An}}$,'' for example, captures the  infected and asymptomatic-nonisolated individuals who do not know whether they are infected. Note that these individuals are infected in actuality, but they do not know that they are infected and they maintain a high contract rate with the rest of the population, being nonisolated. When we perform tests on a segment of a population, the proportion of positive tests is given by the fraction of the infected individuals in the segment relative to the size of the whole segment. For example, using the label of a compartment to also denote the number of individuals in the compartment, since there are $N_{\text{An}} + R_{\text{An}} + I_{\text{An}}$ asymptomatic-nonisolated individuals, if we test $T$  asymptomatic-nonisolated individuals, then the number of positive tests is $
T \times \frac{I_{\text{An}}}{N_{\text{An}} + R_{\text{An}} + I_{\text{An}}}.
$

So far, we indicated two reasons for our model to be optimistic. In addition, our model assumes that close contacts of every positive case are traced and these close contacts remain in isolation for 14 days or until they are tested negative. In other words, our model implicitly assumes that the bottleneck is the testing capacity, not contact tracing capacity. It is important to note that growing the testing capacity in the state beyond a certain level without growing the contact tracing capacity will not be of much benefit, so any decision regarding how much to grow the contact tracing capacity should be directly informed by the plans to grow the testing capacity.

\section{Mathematical Description of the System Dynamics}
\label{section:system-dynamics}

\vspace{10pt}

In this section, we provide a detailed description of the system dynamics of our model.  The state of the system at each time period $t$ is described by a vector 
$$
    \left( {\bm{I}}_{\sf Si}^t, \bm{I}^t_{\sf Ai}, \bm{I}^t_{\sf An}, R^t_{\sf Si}, R^t_{\sf Ai}, R^t_{\sf An}, N^t_{\sf Si}, N^t_{\sf Ai}, N^t_{\sf An}, D^t, \bm{H}^t, \bm{KI}^t, {KR}^t  \right)~,
$$
where the variables $\bm{I}_{\sf Si}^t, \bm{I}^t_{\sf Ai}, \bm{I}^t_{\sf An}, R^t_{\sf Si}, R^t_{\sf Ai}, R^t_{\sf An}, N^t_{\sf Si}, N^t_{\sf Ai}, N^t_{\sf An}$ denote the states associated with the nine compartments on the left side of Figure \ref{fig:comp}. These nine compartments correspond to individuals whose COVID-19 status is unknown.  
The variable $D^t$ denotes the number of individuals who die in period $t$, and the variable $\bm{H}^t$ denotes the number of individuals who are hospitalized in period $t$. The variable $\bm{KI}^t$ captures the number of individuals who have been confirmed to have COVID-19 and are currently infected in period $t$. The variable ${KR}^t$ captures the number of individuals who have been confirmed to have COVID-19 and recovered by period $t$. 

{\bf Description of Sub-compartments:} For our state variables, we use regular font to denote scalars and bold font to denote vectors. We use vector notation when a compartment has multiple sub-compartments representing different subgroups of the population. Here are the descriptions of the sub-compartments.

\begin{itemize}[leftmargin=*]
    \item {Sub-compartments within the $\bm{I}_{\sf Si}$ compartment:}  The $\bm{I}_{\sf Si}$ compartment has 3 sub-compartments: (a) those who will recover naturally without requiring any hospitalization, (b) those who will require hospitalization, and (c) those who will die without access to a COVID-19 diagnostic test. Thus, 
    $$
       \bm{I}^t_{\sf Si} = \left( I^t_{\sf Si}(\mathrm{recovered}), I^t_{\sf Si}(\mathrm{hospitalized}), I^t_{\sf Si}(\mathrm{death}) \right)~.
    $$
    We use the notation  $\bar{I}^t_{\sf Si} = I^t_{\sf Si}(\mathrm{recovered}) \,+\, I^t_{\sf Si}(\mathrm{hospitalized}) \,+\, I^t_{\sf Si}(\mathrm{death})$ to denote the total number of individuals in the $\bm{I}_{\sf Si}$ compartment. The trick we use here is that when an individual is infected, we immediately decide whether this person will recover, will be hospitalized, or will die. We keep the identity of the individual accordingly throughout the simulation.

    \item {Sub-compartments within the $\bm{I}_{\sf Ai}$ compartment:}  The $\bm{I}_{\sf Ai}$ compartment has 2 sub-compartments: (a) those who will never develop COVID-19 symptoms, and (b) those who will show COVID-19 symptoms but are currently pre-symptomatic. Thus,
    $$
        \bm{I}^t_{\sf Ai} = \left( I^t_{\sf Ai}(\mathrm{recovered}), I^t_{\sf Ai}(\mathrm{show ~symptom}) \right)~,
    $$
    and we let $\bar{I}^t_{\sf Ai} ~=~ I^t_{\sf Ai}(\mathrm{recovered}) ~+~ I^t_{\sf Ai}(\mathrm{show ~symptom})$ denote the total number of individuals in the $\bm{I}_{\sf Ai}$ compartment.
    
     \item {Sub-compartments within the $\bm{I}_{\sf An}$ compartments:}  Similar to the $\bm{I}_{\sf Ai}$ compartment,  the $\bm{I}_{\sf An}$ compartment has 2 sub-compartments: (a) those who will never develop COVID-19 symptoms, and (b) those who will show COVID-19 symptoms but are currently pre-symptomatic. Thus,
    $$
         \bm{I}^t_{\sf An} = \left( I^t_{\sf An}(\mathrm{recovered}), I^t_{\sf An}(\mathrm{show ~symptom}) \right)~,
    $$
    and as before, we let $\bar{I}^t_{\sf An} ~=~ I^t_{\sf An}(\mathrm{recovered}) ~+~ I^t_{\sf An}(\mathrm{show ~symptom})$ denote the total number of individuals in the $\bm{I}_{\sf An}$ compartment.

    \item {Sub-compartments within the $\bm{H}$ compartment:}  The $\bm{H}$ compartment consists of two types of individuals: (a) those who will die and (b) those who will eventually recover. Thus, we have
    $\bm{H}^t = \left( H^t(\mathrm{die}), H^t(\mathrm{recovered}) \right)$.
    
    \item {Sub-compartments within the $\bm{KI}$ compartment:}  The $\bm{KI}$ compartment consists of two types of individuals: (a) those who will require hospitalization and (b) those who will recover without visiting a hospital. Therefore,  we have that
    $\bm{KI}^t = \left( {KI}^t(\mathrm{hospitalized}), {KI}^t(\mathrm{recovered}) \right)$.
    
\end{itemize}

{\bf Impact of Testing:} 
We assume that the diagnostic test is 100\% accurate and the result is obtained  instantaneously. However when we perform the test, we cannot differentiate between unknown infected, unknown non-infected, and recovered people.  We can only test people based on their observable characteristics.  At the beginning of each period, we test $T_{\sf Si}$ symptomatic-isolated individuals, $T_{\sf Ai}$ asymptomatic-isolated individuals, and $T_{\sf An}$ asymptomatic-nonisolated individuals.  We assume that the tests are administered at random within each of these populations. Let $\pi_{\sf Si}$, $\pi_{\sf Ai}$, and $\pi_{\sf An}$ denote the fraction of symptomatic isolated, asymptomatic isolated, and  asymptomatic non-isolated individuals, respectively, who receive diagnostic tests.  Then,
\begin{align*}
     \pi_{\sf Si}  &=  \min \left\{ \frac{ T^t_{\sf Si} }{ \displaystyle \bar{I}^t_{\sf Si} ~-~ I^t_{{\sf Si}}(\mathrm{death}) ~+~ R^t_{\sf Si} ~+~ N^t_{\sf Si} } ~,~ 1 \right\},\\
     \pi_{\sf Ai}  &=  \min \left\{ \frac{ T^t_{\sf Ai} }{ \displaystyle \bar{I}^t_{\sf Ai} ~+~ R^t_{\sf Ai} + N^t_{\sf Ai} } ~,~ 1 \right\}, \\
     \pi_{\sf An}  &=  \min \left\{ \frac{ T^t_{\sf An} }{ \displaystyle \bar{I}^t_{\sf An}~+~ R^t_{\sf An} + N^t_{\sf An} }  ~,~ 1 \right\}~.
\end{align*}
The above  fractions drive the dynamics across compartments, which are described below based on the events that can occur in our~model. In the fraction $\pi_{\sf Si}$, we assume that the infected symptomatic-isolated individuals from the death sub-compartment are inaccessible for testing.

{\bf Description of Dynamics Between Compartments:} Our description below is organized by the compartments.  For each compartment, we describe the inflows, outflows, and the new state at time period $t + 1$.  

\begin{itemize}[leftmargin=*]

\item {\bf Infected Symptomatic-Isolated Compartment ($\bm{I}_{\sf Si}$):}  As noted earlier,  the $\bm{I}_{\sf Si}$ compartment has 3 subgroups of individuals: (a) those who will recover naturally without requiring any hospitalization, (b) those who require hospitalization, and (c) those who die without access to a COVID-19 diagnostic test. Thus,
 $$
       \bm{I}^t_{\sf Si} = \left( I^t_{\sf Si}(\mathrm{recovered}), I^t_{\sf Si}(\mathrm{hospitalized}), I^t_{\sf Si}(\mathrm{death}) \right)~,
    $$
and   $\bar{I}^t_{\sf Si} = I^t_{\sf Si}(\mathrm{recovered}) \,+\, I^t_{\sf Si}(\mathrm{hospitalized}) \,+\, I^t_{\sf Si}(\mathrm{death})$ denotes the total number of individuals across the three sub-compartments.


\uline{Inflow}: The inflow to the ${\bm I}_{\sf Si}$ compartment consists of 3 sources given by:
\begin{itemize}[leftmargin=*]
    \item Non-infected symptomatic-isolated ($N_{\sf Si})$ individuals who did not get tested and were newly infected during period $t$. The number of such individuals is given by
\begin{align*}
  \beta_\ell \times \Big(  \bar{I}^t_{\sf An} (1-\pi_{\sf An} )
 + \bar{I}^t_{\sf Ai} \, (1-\pi_{\sf Ai} )  +  \bar{I}^t_{\sf Si} \, (1-\pi_{\sf Si} ) \Big) \times  N^t_{\sf Si}\,(1-\,\pi_{\sf Si} )   ~,
\end{align*}
where $ \bar{I}^t_{\sf An} (1-\pi_{\sf An} )
 + \bar{I}^t_{\sf Ai} \, (1-\pi_{\sf Ai} )  +  \bar{I}^t_{\sf Si} \, (1-\pi_{\sf Si} ) $ represents the infected population who remains untested, and $\beta_\ell$ represents the effective contact rate among isolated individuals. See Section \ref{section:model-parameters} for more details.
 
    \item Infected asymptomatic-nonisolated ($\bm{I}_{\sf An}$) individuals who become symptomatic. The total inflow of such individuals is given by 
$$
    \frac{ I^t_{\sf An}(\mathrm{show~symptom}) }{  {\tt infToSympTime}}~,
$$
    where {\tt infToSympTime} represents the average number of days until an infected person displays symptoms. In our model, we set ${\tt infToSympTime}$ to be $5$ days; see Section \ref{section:model-parameters} for more details.
    
    \item Infected asymptomatic-isolated ($\bm{I}_{\sf Ai}$) individuals who become symptomatic. Using a similar logic as above, the total amount of such inflow is given by $\displaystyle \frac{I^t_{\sf Ai}(\mathrm{show~symptom})}{  {\tt infToSympTime}}$.
\end{itemize}

We assume that a fraction {\tt hospOutOfSympFrac}, currently set at 20\%, of the total inflow to the $\bm{I}_{\sf Si}$ compartment will require hospitalization, and a fraction {\tt deathOutOfSympFrac} (2\%) will die without being tested.  The remaining $1 - {\tt hospOutOfSympFrac} -  {\tt deathOutOfSympFrac} = 78\%$ will recover at home. Justifications for these fractions are given in Section \ref{section:model-parameters}.

\vspace{10pt}
\uline{Outflow}: There are four destinations for the outflow from   the $\bm{I}_{\sf Si}$ compartment.

\begin{itemize}[leftmargin=*]
\item Individuals  in the ``recovered'' sub-compartment, $I^t_{\sf Si}(\mathrm{recovered})$, will move to the recovered asymptomatic-nonisolated ($R_{\sf An}$) compartment at the   rate of $1/{\tt sympToRecoveryTime}$, where {\tt sympToRecoveryTime} represents the average time for a symptomatic individual to recover from the disease. Currently, {\tt sympToRecoveryTime} is  set at 14 days.

\item Individuals in the ``hospitalized'' sub-compartment, $I^t_{\sf Si}(\mathrm{hospitalized})$, will move to the $\bm{H}$ compartment at the rate of $1/{\tt sympToHospTime}$, where ${\tt sympToHospTime}$ (5 days)  represents the average number of days from symptom onset until hospitalization.

\item Individuals in the  ``death'' sub-compartment, $I^t_{\sf Si}(\mathrm{death})$, will move to the death compartment ($D$) at the rate of $1/{\tt sympToDeathTime}$, where {\tt sympToDeathTime} (14 days) denotes the average time from symptoms onset to death.

\item Finally, a fraction $\pi_{\sf Si}$ in every sub-compartment will move to the known infected ($\bm{KI}$)~compartment due to testing.
\end{itemize}

\vspace{10pt}
\uline{Update Equations}: 
Let ${\tt recoverOutOfSympFrac} = 1 - {\tt hospOutOfSympFrac} -  {\tt deathOutOfSympFrac}$.  We have the following update equations.

{\fontsize{10}{10}\selectfont{
\begin{align*}
&I^{t+1}_{\sf Si}(\mathrm{recovered}) \\
&~=~ I^{t}_{\sf Si}(\mathrm{recovered}) \times ( 1 - \pi_{\sf Si}  )   \times \left( 1 - \frac{1}{{\tt sympToRecoveryTime}} \right) \\
&\quad~~~~+~  {\tt recoverOutOfSympFrac}  \times \Bigg[ \beta_\ell \times
 \Big(  \bar{I}^t_{\sf An} (1-\pi_{\sf An} )
 + \bar{I}^t_{\sf Ai} \, (1-\pi_{\sf Ai} )  +  \bar{I}^t_{\sf Si} \, (1-\pi_{\sf Si} ) \Big) \times  N^t_{\sf Si}\,(1-\,\pi_{\sf Si} )  \\
&~~~~~~~~~~~~~~~~~~~~~~~~~~~~~~~~~~~~~~~~~~~~~~~~~ ~+~  \frac{I^t_{\sf An}(\mathrm{show~symptom}) (1-\pi_{\sf An})}{{\tt infToSympTime}}   ~+~ \frac{I^t_{\sf Ai}(\mathrm{show~symptom}) (1 - \pi_{\sf Ai})}{{\tt infToSympTime}}  \Bigg]~, \\
&I^{t+1}_{\sf Si}(\mathrm{hospitalized}) \\
&~=~  I^{t}_{\sf Si}(\mathrm{hospitalized}) \times ( 1 - \pi_{\sf Si}  )   \times \left( 1 - \frac{1}{{\tt sympToHospTime}} \right) \\
&\quad~~~~+~  {\tt hospOutOfSympFrac}  \times \Bigg[  \beta_\ell \times
 \Big(  \bar{I}^t_{\sf An} (1-\pi_{\sf An} )
 + \bar{I}^t_{\sf Ai} \, (1-\pi_{\sf Ai} )  +  \bar{I}^t_{\sf Si} \, (1-\pi_{\sf Si} ) \Big) \times  N^t_{\sf Si}\,(1-\,\pi_{\sf Si} )  \\
&~~~~~~~~~~~~~~~~~~~~~~~~~~~~~~~~~~~~~~~~~~~~~~ ~+~  \frac{I^t_{\sf An}(\mathrm{show~symptom}) (1-\pi_{\sf An})}{{\tt infToSympTime}}   ~+~ \frac{I^t_{\sf Ai}(\mathrm{show~symptom}) (1 - \pi_{\sf Ai})}{{\tt infToSympTime}}  \Bigg]~,\\
&I^{t+1}_{\sf Si}(\mathrm{death}) \\
&~=~  I^{t}_{\sf Si}(\mathrm{death}) \times ( 1 - \pi_{\sf Si}  )   \times \left( 1 - \frac{1}{{\tt sympToDeathTime}} \right) \\
&\quad~~~~+~ {\tt deathOutOfSympFrac} \times \Bigg[  \beta_\ell \times
 \Big(  \bar{I}^t_{\sf An} (1-\pi_{\sf An} )
 + \bar{I}^t_{\sf Ai} \, (1-\pi_{\sf Ai} )  +  \bar{I}^t_{\sf Si} \, (1-\pi_{\sf Si} ) \Big) \times N^t_{\sf Si}\,(1-\,\pi_{\sf Si} )  \\
&~~~~~~~~~~~~~~~~~~~~~~~~~~~~~~~~~~~~~~~~~~~~~~ ~+~  \frac{I^t_{\sf An}(\mathrm{show~symptom}) (1-\pi_{\sf An})}{{\tt infToSympTime}}   ~+~ \frac{I^t_{\sf Ai}(\mathrm{show~symptom}) (1 - \pi_{\sf Ai})}{{\tt infToSympTime}}  \Bigg]~.
\end{align*}
}}

\vspace{10pt}
\item{{\bf Infected Asymptomatic-Isolated Compartment ($\bm{I}_{\sf Ai}$)}:}
We have two sub-compartments with $\bm{I}^t_{\sf Ai} = \left( I^t_{\sf Ai}(\mathrm{recovered}), I^t_{\sf Ai}(\mathrm{show ~symptom}) \right)$, and \mbox{$\bar{I}^t_{\sf Ai} = I^t_{\sf Ai}(\mathrm{recovered}) ~+~ I^t_{\sf Ai}(\mathrm{show ~symptom})$} denotes the total number of individuals across the two sub-compartments.

\vspace{10pt}
\uline{Inflow}: The inflow to the $\bm{I}_{\sf Ai}$ compartment has two sources:
\begin{itemize}[leftmargin=*]
    \item    Non-infected asymptomatic-isolated ($N_{\sf Ai}$) individuals who were not tested and became newly infected during period $t$. The number of such individuals is given by
\begin{align*}
 \beta_\ell \times \Big( \bar{I}^t_{\sf An} (1-\pi_{\sf An} )
 + \bar{I}^t_{\sf Ai} \, (1-\pi_{\sf Ai} )  + \bar{I}^t_{\sf Si} \, (1-\pi_{\sf Si} ) \Big) \times   N^t_{\sf Ai}\,(1-\,\pi_{\sf Ai} )   ~,
\end{align*}
where $\beta_\ell$ denotes the effective contact rate for isolated individuals.

    \item  Infected asymptomatic-nonisolated ($\bm{I}_{\sf An}$) individuals who were not tested but were identified as a close contact of a positive case. These individuals voluntarily self isolate and reduce their contact rate with others. The number of such individuals is given by
 $$
     {\tt contactPerPosCase} \times  \frac{  \bar{I}^t_{\sf Si} \pi_{\sf Si} + \bar{I}^t_{\sf Ai} \pi_{\sf Ai}  + \bar{I}^t_{\sf An} \pi_{\sf An} }{\bar{H}} \times  {\tt likOfBeingInfected} \times  \bar{I}^t_{\sf An}(1-\pi_{\sf An})~,
 $$
where {\tt contactPerPosCase} is the average number of contacts per positive case, currently set at $4$, whereas $\bar{I}^t_{\sf Si} \pi_{\sf Si} + \bar{I}^t_{\sf Ai} \pi_{\sf Ai}  + \bar{I}^t_{\sf An} \pi_{\sf An}$ denotes the number of positive cases identified that time period and $\bar{H}$ is given by
{\fontsize{10}{10}\selectfont{
$$
\bar{H} = \left( {\tt likOfBeingInfected} \times \bar{I}^t_{\sf An}(1-\pi_{\sf An})\right) \,+\, (R^t_{\sf An} ~+~ R^t_{\sf Si} \pi_{\sf Si} ~+~ R^t_{\sf Ai}\pi_{\sf Ai}) \,+\, (N^t_{\sf An} ~+~ N^t_{\sf Si} \pi_{\sf Si} ~+~ N^t_{\sf Ai}\pi_{\sf Ai})~,
$$}}
\!\!\!where the parameter {\tt likOfBeingInfected} measures the likelihood that a contacted individual is in the infected population as compared to the non-infected and recovered populations. We set {\tt likOfBeingInfected} at 5.
\end{itemize}
We assume that a fraction {\tt symptomFrac}, currently set at 50\%, of  the total inflow to the $\bm{I}_{\sf Ai}$ compartment will develop symptoms, and the remaining $1 - {\tt symptomFrac} = 50\%$ will not develop any symptoms. 

\vspace{10pt}
\uline{Outflow}: There are three destinations for the outflow from the $\bm{I}_{\sf Ai}$ compartment.
\begin{itemize}[leftmargin=*]
    \item The individuals  in the ``recovered'' sub-compartment, $I^t_{\sf Ai}(\mathrm{recovered})$, will move to the $R_{\sf An}$ compartment at the rate of $1/{\tt asympToRecoveryTime}$. The parameter {\tt asympToRecoveryTime} represents the average time that an asymptomatic infected person self-isolates after being identified as a close contact. We currently set this value to 10 days.
    
    \item The ``show symptom'' sub-compartment, $I^t_{\sf Ai}(\mathrm{show~symptom})$, will move to the $\bm{I}_{\sf Si}$ compartment at the rate of $1/{\tt infToSympTime}$, where {\tt infToSympTime} (5 days) is the average time until symptom onset.

    \item A fraction $\pi_{\sf Ai}$ of individuals who are tested will move to the known infected ($\bm{KI}$)~compartment.
\end{itemize}

\vspace{10pt}
\uline{Update Equations}: We have the following update equations.
{\fontsize{10}{10}\selectfont{
\begin{align*}
&I^{t+1}_{\sf Ai}(\mathrm{recovered}) \\
&~=~ 
I^{t}_{\sf Ai}(\mathrm{recovered})\times ( 1 - \pi_{\sf Ai}  )   \times \left( 1 - \frac{1}{{\tt asympToRecoveryTime}} \right)
\\
&\quad~~~~~+~  \left( 1 - {\tt symptomFrac} \right)  \times \Bigg[  \beta_\ell \times \Big( \bar{I}^t_{\sf An} (1-\pi_{\sf An} )
 + \bar{I}^t_{\sf Ai} \, (1-\pi_{\sf Ai} )  + \bar{I}^t_{\sf Si} \, (1-\pi_{\sf Si} ) \Big) \times   N^t_{\sf Ai}\,(1-\,\pi_{\sf Ai} )     \\
&~~~~~~~~~~~~~~~ ~+~       {\tt contactPerPosCase} \times  \frac{  \bar{I}^t_{\sf Si} \pi_{\sf Si} + \bar{I}^t_{\sf Ai} \pi_{\sf Ai}  + \bar{I}^t_{\sf An} \pi_{\sf An} }{\bar{H}} \times  {\tt likOfBeingInfected} \times  \bar{I}^t_{\sf An}(1-\pi_{\sf An}) \bigg]~, \\
&I^{t+1}_{\sf Ai}(\mathrm{show~symptom}) \\
&~=~  I^{t}_{\sf Ai}(\mathrm{show~symptom}) \times ( 1 - \pi_{\sf Ai}  )   \times \left( 1 - \frac{1}{{\tt infToSympTime}} \right) \\
&\quad~~~~~+~  {\tt symptomFrac}   \times \Bigg[  \beta_\ell \times  \Big( \bar{I}^t_{\sf An} (1-\pi_{\sf An} )
 + \bar{I}^t_{\sf Ai} \, (1-\pi_{\sf Ai} )  + \bar{I}^t_{\sf Si} \, (1-\pi_{\sf Si} ) \Big) \times  N^t_{\sf Ai}\,(1-\,\pi_{\sf Ai} )     \\
&~~~~~~~~~~~~~~~ ~+~       {\tt contactPerPosCase} \times  \frac{  \bar{I}^t_{\sf Si} \pi_{\sf Si} + \bar{I}^t_{\sf Ai} \pi_{\sf Ai}  + \bar{I}^t_{\sf An} \pi_{\sf An} }{\bar{H}} \times  {\tt likOfBeingInfected} \times  \bar{I}^t_{\sf An}(1-\pi_{\sf An}) \bigg]~.
\end{align*}
}}


\vspace{10pt}
\item{{\bf Infected Asymptomatic-Nonisolated Compartment ($\bm{I}_{\sf An}$)}:}
We have two sub-compartments, $\bm{I}^t_{\sf An} = \left( I^t_{\sf An}(\mathrm{recovered}), I^t_{\sf An}(\mathrm{show ~symptom}) \right)$, and we denote the total number of individuals across the two sub-compartments as $\bar{I}^t_{\sf An} = I^t_{\sf An}(\mathrm{recovered}) ~+~ I^t_{\sf An}(\mathrm{show ~symptom})$.  

\vspace{10pt}
\uline{Inflow}: The inflow to the $\bm{I}_{\sf An}$ compartment comes from non-infected asymptomatic-nonisolated $N_{\sf An}$ individuals who become newly infected during period $t$. The number of such individuals is given by
\begin{align*}
 \Big( \beta_h \bar{I}^t_{\sf An} (1-\pi_{\sf An} )
 + \beta_\ell \bar{I}^t_{\sf Ai} \, (1-\pi_{\sf Ai} )  + \beta_\ell \bar{I}^t_{\sf Si} \, (1-\pi_{\sf Si} ) \Big) \times N^t_{\sf An}\,(1-\,\pi_{\sf An} )   ~,
\end{align*}
where $\beta_h$ represents the effective contact rate among nonisolated individuals. See Section \ref{section:model-parameters} for more details. We assume that a fraction {\tt symptomFrac}, currently set at 50\%, of  the total inflow to the $\bm{I}_{\sf An}$ compartment will develop symptom, and the remaining $1 - {\tt symptomFrac} = 50\%$ will not develop any symptoms. 


\vspace{10pt}
\uline{Outflow}: The outflow from the $\bm{I}_{\sf An}$ compartment has four destinations.
\begin{itemize}[leftmargin=*]
    \item The individuals  in the ``recovered'' sub-compartment, $I^t_{\sf An}(\mathrm{recovered})$, will move to the $R_{\sf An}$ compartment at the rate of $1/{\tt asympToRecoveryTime}$.
    
    \item The individuals  in the ``show symptom'' sub-compartment, $I^t_{\sf An}(\mathrm{show~symptom})$, will move to the $\bm{I}_{\sf Si}$ compartment at the rate of $1/{\tt infToSympTime}$.
    
    \item   Some individuals who were not tested will reduce their contacts and move to the $\bm{I}_{\sf Ai}$ compartment. The number of such individuals is
$$
     {\tt contactPerPosCase} \times  \frac{  \bar{I}^t_{\sf Si} \pi_{\sf Si} + \bar{I}^t_{\sf Ai} \pi_{\sf Ai}  + \bar{I}^t_{\sf An} \pi_{\sf An} }{\bar{H}} \times  {\tt likOfBeingInfected} \times  \bar{I}^t_{\sf An}(1-\pi_{\sf An})~.
 $$

    \item A fraction $\pi_{\sf An}$ of individuals are tested and moved to the known infected $\bm{KI}$ compartment.
\end{itemize}

\vspace{10pt}
\uline{Update Equations}: We have the following update equations.
{\fontsize{10}{10}\selectfont{
\begin{align*}
&I^{t+1}_{\sf An}(\mathrm{recovered}) \\
&~=~ I^{t}_{\sf An}(\mathrm{recovered}) \times ( 1 - \pi_{\sf An})   \times  \bigg( 1 - \frac{1}{{\tt asympToRecoveryTime}}  \\
&~~~~~~~~~~~~~~~~~~~~~~~~~~~~~~~ ~-~   {\tt contactPerPosCase} \times  \frac{  \bar{I}^t_{\sf Si} \pi_{\sf Si} + \bar{I}^t_{\sf Ai} \pi_{\sf Ai}  + \bar{I}^t_{\sf An} \pi_{\sf An} }{\bar{H}} \times  {\tt likOfBeingInfected} \bigg)\\
&\quad~~~~+~  (1 ~-~ {\tt symptomFrac}) \times \bigg[   \Big( \beta_h \bar{I}^t_{\sf An} (1-\pi_{\sf An} )
 + \beta_\ell \bar{I}^t_{\sf Ai} \, (1-\pi_{\sf Ai} )  + \beta_\ell \bar{I}^t_{\sf Si} \, (1-\pi_{\sf Si} ) \Big) \times N^t_{\sf An}\,(1-\,\pi_{\sf An} )   \bigg]~, \\
&I^{t+1}_{\sf An}(\mathrm{show~symptom}) \\
&~=~  I^{t}_{\sf An}(\mathrm{show~symptom}) \times ( 1 - \pi_{\sf An}  )   \times \bigg( 1 - \frac{1}{{\tt infToSympTime}} \\
&~~~~~~~~~~~~~~~~~~~~~~~~~~~~~~~ ~-~   {\tt contactPerPosCase} \times  \frac{  \bar{I}^t_{\sf Si} \pi_{\sf Si} + \bar{I}^t_{\sf Ai} \pi_{\sf Ai}  + \bar{I}^t_{\sf An} \pi_{\sf An} }{\bar{H}} \times  {\tt likOfBeingInfected} \bigg)\\
&\quad~~~~+~  {\tt symptomFrac} \times \bigg[   \Big( \beta_h \bar{I}^t_{\sf An} (1-\pi_{\sf An} )
 + \beta_\ell \bar{I}^t_{\sf Ai} \, (1-\pi_{\sf Ai} )  + \beta_\ell \bar{I}^t_{\sf Si} \, (1-\pi_{\sf Si} ) \Big) \times N^t_{\sf An}\,(1-\,\pi_{\sf An} )   \bigg]~.
\end{align*}
}}


\vspace{10pt}
\item{{\bf Non-infected Symptomatic-Isolated Compartment ($N_{\sf Si}$)}:} 

\vspace{10pt}
\uline{Inflow}: The inflow to the $N_{\sf Si}$ compartment has two sources:
\begin{itemize}[leftmargin=*]
    \item Non-infected asymptomatic-nonisolated ($N_{\sf An}$) individuals who did not get tested and developed symptoms caused by another condition or disease such as the seasonal flu. There are $N^t_{\sf An} ~+~ N^t_{\sf Si} \pi_{\sf Si} ~+~ N^t_{\sf Ai}\pi_{\sf Ai}$ such individuals. The rate that an individual develops symptoms due to a non-COVID-19 disease is given by the parameter {\tt nonCOVIDSymptRate}, which we currently set at 1/1200; see Section \ref{section:model-parameters} for more details on how we arrive at this number.
    
    \item Non-infected asymptomatic-isolated ($N_{\sf Ai})$ individuals can also develop flu-like symptoms. There are $N^{t}_{\sf Ai} \times ( 1 - \pi_{\sf Ai})$ such individuals.
\end{itemize}

\vspace{10pt}
\uline{Outflow}: The outflow from the $N_{\sf Si}$ compartment has three destinations:
\begin{itemize}[leftmargin=*]
    \item Individuals  who are tested will return negative and move to the $N_{\sf An}$ compartment. This causes them to increase their contact with other individuals.
    \item Some individuals will become infected from coming in contact with infected people.
    \item Individuals leave self-quarantine at a rate of $1/{\tt selfQuarTime}$, where {\tt selfQuarTime} is the average time that an individual self-isolates. We currently set this value to 10 days.
\end{itemize}

\vspace{10pt}
\uline{Update Equations}: We have the following update equations.
{\fontsize{10}{10}\selectfont{
\begin{align*}
N^{t+1}_{\sf Si} &~=~ N^{t}_{\sf Si} \times ( 1 - \pi_{\sf Si})   \times \left( 1 ~-~ \dfrac{1}{{\tt selfQuarTime}} ~-~ \beta_\ell \times \Big(  \bar{I}^t_{\sf An} (1-\pi_{\sf An} )
 + \bar{I}^t_{\sf Ai} \, (1-\pi_{\sf Ai} )  +  \bar{I}^t_{\sf Si} \, (1-\pi_{\sf Si} ) \Big)  \right) \\
 &\quad~~~~+~  {\tt nonCOVIDSymptRate} \times \left(  N^t_{\sf Si} \pi_{\sf Si} ~+~ N^t_{\sf Ai} ~+~ N^t_{\sf An} \right)~. 
\end{align*}
}}

\vspace{10pt}
\item{{\bf Non-infected Asymptomatic-Isolated Compartment ($N_{\sf Ai}$)}:}

\vspace{10pt}
\uline{Inflow}: The inflow to the $N_{\sf Ai}$ compartment comes from individuals in the $N_{\sf An}$ compartment who reduce their exposure to other individuals after being identified as a close contact. The number of such individuals is given~by
$$
    {\tt contactPerPosCase} \times  \frac{  \bar{I}^t_{\sf Si} \pi_{\sf Si} + \bar{I}^t_{\sf Ai} \pi_{\sf Ai}  + \bar{I}^t_{\sf An} \pi_{\sf An} }{\bar{H}} \times  \left( N^t_{\sf An} ~+~ N^t_{\sf Si} \pi_{\sf Si} ~+~ N^t_{\sf Ai} \pi_{\sf Ai} \right)~,
$$
where {\tt contactPerPosCase} is the average number of contacts per positive case, currently set at~$4$, and  $\bar{H}$ denotes the total number of high-contact individuals; see the discussion in the $\bm{I}_{\sf Ai}$ compartment.

\vspace{10pt}
\uline{Outflow}: The outflow from the $N_{\sf Ai}$ compartment has four destinations.
\begin{itemize}[leftmargin=*]
\item Individuals who are tested will test negative and move to the $N_{\sf An}$ compartment, end their self-isolation, and increase their contacts.
\item Untested individuals leave self-isolation at a rate of $1/{\tt selfQuarTime}$.
\item Some individuals will become infected and move to the $\bm{I}_{\sf Ai}$ compartment.  
\item Some individuals will develop flu-like symptoms but not COVID-19, at the rate of  {\tt nonCOVIDSymptRate}.
\end{itemize}

\vspace{10pt}
\uline{Update Equations}: We have the following update equations.
\begin{align*}
N^{t+1}_{\sf Ai} &~=~ N^{t}_{\sf Ai} \times ( 1 - \pi_{\sf Ai})   \times \bigg( 1 ~-~ \frac{1}{{\tt selfQuarTime}} ~-~ {\tt nonCOVIDSymptRate} \\
&~~~~~~~~~~~~~~~~~~~~~~~~~~~~~~~~~ ~-~  \beta_\ell \times \Big( \bar{I}^t_{\sf An} (1-\pi_{\sf An} )
 + \bar{I}^t_{\sf Ai} \, (1-\pi_{\sf Ai} )  + \bar{I}^t_{\sf Si} \, (1-\pi_{\sf Si} ) \Big)  \bigg) \\
 &\quad~~~~~+~       {\tt contactPerPosCase} \times  \frac{  \bar{I}^t_{\sf Si} \pi_{\sf Si} + \bar{I}^t_{\sf Ai} \pi_{\sf Ai}  + \bar{I}^t_{\sf An} \pi_{\sf An} }{\bar{H}} \times  \left( N^t_{\sf An} ~+~ N^t_{\sf Si} \pi_{\sf Si} ~+~ N^t_{\sf Ai} \pi_{\sf Ai} \right)~.
\end{align*}


\vspace{10pt}
\item{{\bf Non-infected Asymptomatic-Nonisolated Compartment ($N_{\sf An}$)}:}

\vspace{10pt}
\uline{Inflow}: The inflow to the $N_{\sf An}$ compartment comes from three sources. 
First, individuals from the $N_{\sf Si}$ compartment recover from non-COVID-19 symptoms at a rate of $\dfrac1{\tt selfQuarTime}$. 
Second, individuals from the $N_{\sf Ai}$ compartment leave self-isolation and increase their contact with others at a rate of ${\dfrac1{\tt selfQuarTime}}$.   
Third, individuals from the $N_{\sf Si}$ and $N_{\sf Ai}$ compartments who test negative will also increase their contact rates.

\vspace{10pt}
\uline{Outflow}: The outflow from the $N_{\sf An}$ compartment has three destinations.  First, some individuals are identified as close contacts and move to the $N_{\sf Ai}$ compartment, thereby reducing their contact levels. Second, other individuals will develop flu-like symptoms independent of COVID and move to the $\bm{N}_{\sf Si}$ compartment at the rate of {\tt nonCOVIDSymptRate}. Third,  some individuals will become infected and move to the $\bm{I}_{\sf An}$ compartment.

\vspace{10pt}
\uline{Update Equations}: We have the following update equations.
\begin{align*}
N^{t+1}_{\sf An} &~=~ \left( N^t_{\sf An} ~+~ N^t_{\sf Si} \pi_{\sf Si} ~+~ N^t_{\sf Ai}\pi_{\sf Ai} \right)  \\ 
&~~~~~~~~~~~~ \times \Bigg( 1 ~-~ {\tt nonCOVIDSymptRate}  ~-~   {\tt contactPerPosCase} \times  \frac{  \bar{I}^t_{\sf Si} \pi_{\sf Si} + \bar{I}^t_{\sf Ai} \pi_{\sf Ai}  + \bar{I}^t_{\sf An} \pi_{\sf An} }{\bar{H}}   \\
& ~~~~~~~~~~~~~~~~~~~~ ~-~  \Big( \beta_h \bar{I}^t_{\sf An} (1-\pi_{\sf An} )
 + \beta_\ell \bar{I}^t_{\sf Ai} \, (1-\pi_{\sf Ai} )  + \beta_\ell \bar{I}^t_{\sf Si} \, (1-\pi_{\sf Si} ) \Big) 
\Bigg) \\
 &\quad~~~~~~~~~+~       \frac{N_{\sf Ai}^t (1-\pi_{\sf Ai} )}{{\tt selfQuarTime}}  ~+~   \frac{ N_{\sf Si}^t (1-\pi_{\sf Si}) }{{\tt selfQuarTime}} ~.
\end{align*}


\vspace{10pt}
\item{{\bf Recovered Symptomatic-Isolated Compartment ($R_{\sf Si}$)}:}

\vspace{10pt}
\uline{Inflow}: The inflow to the $R_{\sf Si}$ compartment has two sources. Recovered asymptomatic-nonisolated ($R_{\sf An}$) and recovered asymptomatic-isolated ($R_{\sf Ai}$) individuals, who did not get tested, can develop flu-like symptoms independent of COVID-19, at the rate of {\tt nonCOVIDSymptRate}.

\vspace{10pt}
\uline{Outflow}: The outflow from the $R_{\sf Si}$ compartment occurs from  individuals who recover from non-COVID-19 symptoms at the rate of $1/{\tt selfQuarTime}$.

\vspace{10pt}
\uline{Update Equations}: We have the following update equations.
{\fontsize{10}{10}\selectfont{
\begin{align*}
R^{t+1}_{\sf Si} &~=~ R^{t}_{\sf Si} \times ( 1 - \pi_{\sf Si})   \times \left( 1 ~-~ \frac{1}{{\tt selfQuarTime}} \right)  ~+~ {\tt nonCOVIDSymptRate} \times \left(  R^t_{\sf Si} \pi_{\sf Si} ~+~ R^t_{\sf Ai} ~+~ R^t_{\sf An} \right) ~.
\end{align*}
}}


\vspace{10pt}
\item{{\bf Recovered Asymptomatic-Isolated Compartment ($R_{\sf Ai}$)}:}

\vspace{10pt}
\uline{Inflow}: The inflow to the $R_{\sf Ai}$ compartment  comes from individuals in the $R_{\sf An}$ compartment who reduce their exposure to other individuals after being identified as a close contact.    The number of such individuals is given by
$$
    {\tt contactPerPosCase} \times  \frac{  \bar{I}^t_{\sf Si} \pi_{\sf Si} + \bar{I}^t_{\sf Ai} \pi_{\sf Ai}  + \bar{I}^t_{\sf An} \pi_{\sf An} }{\bar{H}} \times  \left( R^t_{\sf An} ~+~ R^t_{\sf Si} \pi_{\sf Si} ~+~ R^t_{\sf Ai} \pi_{\sf Ai} \right)~,
$$

\vspace{10pt}
\uline{Outflow}: The outflow from the $R_{\sf Ai}$ compartment has two destinations. First, untested individuals leave isolation at the rate of $1/{\tt selfQuarTime}$. Second, other individuals will develop symptoms independent of COVID-19, at the rate of {\tt nonCOVIDSymptRate}.

\vspace{10pt}
\uline{Update Equations}: We have the following update equations.
\begin{align*}
R^{t+1}_{\sf Ai} &~=~ R^{t}_{\sf Ai} \times ( 1 - \pi_{\sf Ai})   \times \left( 1 ~-~ \frac{1}{{\tt selfQuarTime}} ~-~  {\tt nonCOVIDSymptRate}  \right) \\
 &\quad~~~~~+~     {\tt contactPerPosCase} \times  \frac{  \bar{I}^t_{\sf Si} \pi_{\sf Si} + \bar{I}^t_{\sf Ai} \pi_{\sf Ai}  + \bar{I}^t_{\sf An} \pi_{\sf An} }{\bar{H}} \times  \left( R^t_{\sf An} ~+~ R^t_{\sf Si} \pi_{\sf Si} ~+~ R^t_{\sf Ai} \pi_{\sf Ai} \right)~.
\end{align*}


\vspace{10pt}
\item{{\bf Recovered Asymptomatic-Nonisolated Compartment ($R_{\sf An}$)}:}

\vspace{10pt}
\uline{Inflow}: The inflow to the $R_{\sf An}$ compartment has six sources:
\begin{itemize}
    \item  Untested people from the $\bm{I}_{\sf Si}$ compartment who recovered naturally.
    \item  Untested people from the $\bm{I}_{\sf Ai}$ compartment who recovered naturally.
    \item  Untested people from the $\bm{I}_{\sf An}$ compartment who recovered naturally.
        \item People from the $R_{\sf Ai}$ and $R_{\sf Si}$ compartments who test negative.
    \item  People from the $R_{\sf Ai}$ compartment who leave self-quarantine.
    \item People from the $R_{\sf Si}$ compartment whose non-COVID-19 symptoms disappeared.
\end{itemize}

\vspace{10pt}
\uline{Outflow}: The outflow from the $R_{\sf An}$ compartment occurs in two ways.  First,
some individuals are identified as close contacts and move to the $R_{\sf Ai}$ compartment. Second, some individuals will develop symptoms independent of COVID-19, at the rate of {\tt nonCOVIDSymptRate}.

\vspace{10pt}
\uline{Update Equations}: We have the following update equations.
{\fontsize{10}{10}\selectfont{
\begin{align*}
R^{t+1}_{\sf An} &~=~ \left( R^t_{\sf An} ~+~ R^t_{\sf Si} \pi_{\sf Si} ~+~ R^t_{\sf Ai} \pi_{\sf Ai} \right)  \\ &\quad~~~~~~~~~~~~~ \times \Bigg( 1 ~-~  {\tt nonCOVIDSymptRate}   ~-~  {\tt contactPerPosCase} \times  \frac{  \bar{I}^t_{\sf Si} \pi_{\sf Si} + \bar{I}^t_{\sf Ai} \pi_{\sf Ai}  + \bar{I}^t_{\sf An} \pi_{\sf An} }{\bar{H}}   \Bigg) \\
&\quad~~~~ 
~+~ \frac{R_{\sf Ai}^t (1-\pi_{\sf Ai})}{{\tt selfQuarTime}}    
~+~ \frac{R_{\sf Si}^t (1-\pi_{\sf Si})}{{\tt selfQuarTime}}   
~+~ \frac{ \bar{I}^t_{\sf Si} (1 - \pi_{\sf Si})}{{\tt sympToRecoveryTime}}\\
&\quad~~~~ 
~+~ \frac{ \bar{I}^t_{\sf Ai} (1 - \pi_{\sf Ai})}{{\tt asympToRecoveryTime}} 
~+~ \frac{ \bar{I}^t_{\sf An} (1 - \pi_{\sf An})}{{\tt asympToRecoveryTime}}~.
\end{align*}
}}


\vspace{10pt}
\item{{\bf Known Infected Compartment ($\bm{KI}$)}:}
This compartment has two sub-compartments, with    $\bm{KI}^t = \left( {KI}^t(\mathrm{hospitalized}), H^t(\mathrm{recovered}) \right)$ and we have the following dynamics.

\vspace{10pt}
\uline{Inflow}: There are three sources of inflow. The infected individuals in $\bm{I}_{\sf Si}$, $\bm{I}_{\sf Ai}$, and $\bm{I}_{\sf An}$ compartments who are tested. We assume that individuals from the $\bm{I}_{\sf Si}(\mathrm{hospitalized})$ sub-compartment flow into the ${KI}^t(\mathrm{hospitalized})$ sub-compartment, and that individuals from the $\bm{I}_{\sf Si}(\mathrm{recovered})$ sub-compartment flow into the ${KI}^t(\mathrm{recovered})$ sub-compartment. We assume that {\tt hospOutOfSymptFrac} (20\%) of the individuals from the $\bm{I}_{\sf Ai}$ and $\bm{I}_{\sf An}$ compartments require hospitalization, while the remaining 1 - {\tt hospOutOfSymptFrac} = 80\% will recover naturally.

\vspace{10pt}
\uline{Outflow}: Individuals who require hospitalization move to the $\bm{H}$ compartment at the rate of $1/{\tt infToHospTime}$, where {\tt infToHospTime} represents the average time until an infected person requires hospitalization. We currently set this to 5 days.  Individuals who will recover move to the $KR$ compartment at the rate of
$1/{\tt sympToRecoveryTime}$, where {\tt sympToRecoveryTime} denotes the average recovery time. We set this to 14 days.

\vspace{10pt}

\uline{Update Equations}: Let ${\tt recoverOutOfSymptFrac} = 1 - {\tt hospOutOfSymptFrac} = 80\%$. Then,
{\fontsize{10}{10}\selectfont{
\begin{align*}
{KI}^{t+1}(\mathrm{recovered}) &~=~ \bigg({KI}^t(\mathrm{recovered}) ~+~  I^t_{\sf Si}(\textrm{recovered}) \pi_{\sf Si} ~+~ I^t_{\sf Ai}(\textrm{recovered}) \pi_{\sf Ai} ~+~ I^t_{\sf An}(\textrm{recovered})\pi_{\sf An}\\
&~~~~~~~~~~~~~~~~~ +~{\tt recoverOutOfSymptFrac}  \times  \left[ I^t_{\sf Ai}(\mathrm{sympt.}) \pi_{\sf Ai} ~+~  I^t_{\sf An}(\mathrm{sympt.})\pi_{\sf An} \right]  \bigg)\\
&~~~~~~~~~~~~~~~~~~~~~~~~~~\times \left( 1 -  \frac{1}{{\tt sympToRecoveryTime}}  \right)~, \\
{KI}^{t+1}(\mathrm{hospitalized}) &~=~ \bigg(  {KI}^t(\mathrm{hospitalized})  ~~+~   I^t_{\sf Si}(\mathrm{hospitalized}) \pi_{\sf Si}\\
&~~~~~~~~~~~~~~~~~ + {\tt hospOutOfSymptFrac} \times \left[ I^t_{\sf Ai}(\mathrm{sympt.}) \pi_{\sf Ai} \,+\,  I^t_{\sf An}(\mathrm{sympt.})\pi_{\sf An}   \right] \bigg)\\
&~~~~~~~~~~~~~~~~~~~~~~~~~~\times \left( 1 -  \frac{1}{{\tt infToHospTime}}  \right)~.\\
\end{align*}
}}



\vspace{-5pt}
\item{{\bf Hospitalization Compartment ($\bm{H}$)}:}
This compartment has two sub-compartments, with    $\bm{H}^t = \left( {H}^t(\mathrm{die}), H^t(\mathrm{recovered}) \right)$.

\vspace{10pt}
\uline{Inflow}: There are two sources of inflow to the hospitalization compartment: the infected individuals from the $\bm{I}_{\sf Si}$ and $\bm{KI}$ compartments who require hospitalization. We assume that a fraction ${\tt deathFrac}$, currently set at 1/3, of the inflow to the hospitalization compartment will die. We assume the remaining $1- {\tt deathFrac} = 2/3$ will recover.

\vspace{10pt}
\uline{Outflow}: Individuals who die leave the hospital at the rate of $1/{\tt hospToDeathTime}$, where {\tt hospToDeathTime} is the average time between hospitalization and death.
Individuals who will recover move to the $KR$ compartment at the rate of $1/{\tt hospToRecoveryTime}$, where {\tt hospToRecoveryTime} is the average time between hospitalization and recovery. We currently set both to 14 days.

\vspace{10pt}
\uline{Update Equations}: Let ${\tt recFrac} = 1- {\tt deathFrac} = 2/3$.
We have the following update equations.
{\fontsize{10}{10}\selectfont{
\begin{align*}
H^{t+1}(\mathrm{die}) &~=~ H^{t}(\mathrm{die}) \times \left( 1 -  \frac{1}{{\tt hospToDeathTime}}  \right) ~+~ \\
&\quad \quad \frac{ {\tt deathFrac} }{{\tt infToHospTime}}  \times \Bigg(  I^t_{\sf Si}(\mathrm{hosp.})\times (1-\pi_{\sf Si}) ~+~ 
{KI}^t(\mathrm{hosp.}) ~+~ 
I^t_{\sf Si}(\mathrm{hosp.}) \pi_{\sf Si} \\
&\quad \quad ~~~~~~~~~~~~~~~~~~~~~~~  ~+~ {\tt hospOutOfSymptFrac} \times \left[ I^t_{\sf Ai}(\mathrm{sympt.}) \pi_{\sf Ai} ~+~  I^t_{\sf An}(\mathrm{sympt.}) \pi_{\sf An}   \right]  \Bigg)~, \\
H^{t+1}(\mathrm{recovered}) &~=~ H^{t}(\mathrm{recovered}) \times \left( 1 -  \frac{1}{{\tt hospToRecoveryTime}}  \right) ~+~ \\
&\quad \quad \frac{ {\tt recFrac} }{{\tt infToHospTime}}  \times \Bigg(  I^t_{\sf Si}(\mathrm{hosp.})\times (1-\pi_{\sf Si}) ~+~ 
{KI}^t(\mathrm{hosp.}) ~+~ 
I^t_{\sf Si}(\mathrm{hosp.}) \pi_{\sf Si} \\
&\quad \quad ~~~~~~~~~~~~~~~~~~~~~~~ ~+~ {\tt hospOutOfSymptFrac} \times \left[ I^t_{\sf Ai}(\mathrm{sympt.}) \pi_{\sf Ai} ~+~  I^t_{\sf An}(\mathrm{sympt.}) \pi_{\sf An}   \right]  \Bigg)~.
\end{align*}
}}


\vspace{-10pt}
\item{{\bf Known Recovery Compartment ($KR$)}:}

\vspace{10pt}
\uline{Update Equations}: Let ${\tt recoverOutOfSymptFrac} = 1 - {\tt hospOutOfSymptFrac} = 80\%$. We have the following update equations.
\begin{align*}
{KR}^{t+1} ~=~ {KR}^{t} &~+~  \frac{H^{t}(\mathrm{rec.}) }{{\tt hospToRecoveryTime}}  ~+~
\frac{ {KI}^t(\mathrm{rec.}) }{{\tt sympToRecoveryTime}}\\
&~+~
\frac{   I^t_{\sf Si}(\textrm{rec.}) \pi_{\sf Si} ~+~ I^t_{\sf Ai}(\textrm{rec.}) \pi_{\sf Ai} ~+~ I^t_{\sf An}(\textrm{rec.})\pi_{\sf An}}{{\tt sympToRecoveryTime}} \\
&~+~ \frac{ {\tt recoverOutOfSymptFrac} \times \Big[ I^t_{\sf Ai}(\mathrm{sympt.}) \pi_{\sf Ai} ~+~ I^t_{\sf An}(\mathrm{sympt.})\pi_{\sf An} \Big] }{ {\tt sympToRecoveryTime} }~.
\end{align*}



\vspace{10pt}
\item{{\bf Death Compartment ({\sf D})}:}

\vspace{10pt}
\uline{Update Equations}: We have the following update equations.
\begin{align*}
{D}^{t+1} &~=~ {D}^{t} ~+~ \frac{ H^{t}(\mathrm{die}) }{{\tt hospToDeathTime}} + \dfrac{I^t_{\sf Si}(\mathrm{death})}{{\tt sympToDeathTime}}~.
\end{align*}

\end{itemize}

\section{Model Parameters and Their Estimated Values}
\label{section:model-parameters}


\vspace{10pt}
In this section, we discuss how we obtain the estimates for different model parameters.

\noindent \textbf{Estimating Flow Rates Between Compartments:} 

\noindent We assume an incubation period of $5$ days~\cite{Linton2020}. 
After an individual develops symptoms, we assume it takes another $5$ days for the individual to become hospitalized~\cite{Wang2020}. Specifically, we set ${\tt infToSympTime}={\tt infToHospTime} = {\tt sympToHospTime}=5$.

We set the recovery time for symptomatic individuals to be $14$ days. In our model, we do not differentiate between recovery at home versus at the hospital and set ${\tt sympToRecoveryTime} = {\tt hospToRecoveryTime} =14$. We set the time for a symptomatic individual to die to be $14$ days. Similar to the recovery time, we do not differentiate between deaths at home versus at the hospital and set both ${\tt sympToDeathTime}={\tt hospToDeathTime}=14$~\cite{Linton2020}.  

We assume that individuals who self-isolate after being identified as a close-contact or develop symptoms from a non-COVID-19 illness spend an average of $10$ days reducing their contact with others. While the WHO recommends a quarantine period of 14 days, the CDC gives a less stringent recommendation of 10 days~\cite{CDC2020}. We set ${\tt selfQuarTime}={\tt asympToRecoveryTime}=10$. 

In our model, non-infected and recovered individuals develop symptoms due to non-COVID-19 diseases at a rate of ${\tt nonCOVIDSymptRate}=\frac1{1200}$. This rate is based on the assumption that $10\%$ of the population is infected over the course of $4$ months. This is roughly estimated using the flu symptom rate in past years as reported by the CDC~\cite{CDCFlu2020}. We set $N=10,490,000$, the population of North Carolina in 2020. We also assume that $34,966$ individuals, corresponding to $0.3\%$ of the population, initially has symptoms due to the seasonal flu. This is calculated assuming that flu symptoms last for an average of $4$ days.

\noindent \textbf{Estimating Disease Dynamics:} 

\noindent We use ${\tt sympOutOfInfFrac}$ to denote the fraction of COVID-19 infected individuals who develop symptoms. Numerous reports have given different estimates for this fraction: from $82\%$ on the Diamond Princess cruise ship~\cite{MKZC2020} to roughly $40\%$ on the USS Theodore Roosevelt~\cite{Reuters2020}. We set ${\tt sympOutOfInfFrac}=50\%$.

The parameter ${\tt hospOutOfSympFrac}$ denotes the fraction of symptomatic individuals who need hospitalization. We set this value to be $20\%$. The parameter ${\tt deathOutOfHospFrac}$ denotes the fraction of hospitalized individuals who die. We set this value to be $33.\bar{3}\%$. Both of these values are roughly estimated using historical hospitalization and death numbers as reported by New York City \cite{NYCdata}. Finally, we assume that some portion of the infected population die at home without hospitalization. We denote this portion as ${\tt deathOutOfInfFrac}$ and set it to $2\%$. We estimate this value using the number of probable deaths as reported by the City.

\noindent\textbf{Infection Dynamics and Estimating Contact Rates:}

\noindent The basic reproductive number, $R_0$, captures the average number of new infections produced by each infected individual. There are numerous recent studies estimating the value of $R_0$ for COVID-19. These estimates vary widely, from anywhere between 2.2~\cite{Li2020} and 5.7~\cite{Sanche2020}. Furthermore, it is possible that $R_0$ varies across different countries and cities due to climate, environmental, and sociological differences. As a result, we set $R_0=3.04$ according to a tuning procedure using historical hospitalization and death statistics as reported by North Carolina. We briefly discuss the details of this procedure in Section \ref{section:model-validation}.

The rate of new infections is controlled by $\beta$, the contact rate between infected and non-infected populations. In a standard compartmentalized model, the number of new infections during any time period is captured by 
$$\beta \times \{\# \text{ of non-infected individuals}\} \times \{\# \text{ of infected individuals}\}.$$
In our model, we have a similar term for every pair of infected and non-infected compartments. We employ two different contact rates $\beta_h$ and $\beta_\ell$, so we differentiate their usage according to whether we are considering isolating or non-isolating compartments.

We set $\beta_h=\frac{R_0}{N\times {\tt sympToRecoveryTime}}$ and use this as the contact rate between infected and \mbox{non-infected} compartments that are both non-isolating.
When at least one of the compartments is isolating or symptomatic, we assume that contact between these populations is reduced by one third and use the contact rate $\beta_\ell = \frac23\times \frac{R_0 }{N \times {\tt sympToRecoveryTime}}$.

Beginning on March 27, we reduce both $\beta_h$ and $\beta_\ell$ by a factor of $1/2$. This reduction is due to social distancing measures associated with the NC stay-at-home order. We set this reduction to $1/2$ according to data from IHME that reports that starting April 1, mobility in North Carolina was reduced by approximately 50\%~\cite{ihme2020}. On May 8, we assume that Phase 1 relaxes social distancing to 25\% of the stay-at-home level. On May 22, we assume that Phase 2 relaxes social distancing to 50\% of the stay-at-home level.

\noindent\textbf{Dynamics of Contact Tracing:}

\noindent The parameter ${\tt contactPerPosCase}$ denotes the average number of close contacts that are identified via tracing for each COVID-19 positive individual. We assume that close contacts are always identified from within the non-isolating population. To estimate ${\tt contactPerPosCase}$ and the proportion of infected and non-infected individuals within the close-contact group, we use the results of~\cite{Bi2020}, a comprehensive study on contact tracing. This study reports the results of contact tracing in Shenzhen, China but the index cases are mostly travelers arriving from Hubei province. (To the best of our knowledge there are no detailed studies on contact tracing in the United States. Therefore, we use findings reported for the Hubei province in China and its capital Wuhan to estimate ${\tt contactPerPosCase}$ and to get a rough approximation for ${\tt likOfBeingInfected}$ as explained in the following.)

In this study, the authors identified 1,286 close contacts based on 292 COVID-positive cases. This suggests an average close contact group size of $1286 / 292 = 4.4$ for each COVID-positive individual identified. According to recent news articles, the state of Massachusetts, which has recently initiated a comprehensive contact-tracing effort, found that the average close contact group size is two~\cite{nbcb2020}. Therefore, we set the parameter ${\tt contactPerPosCase}$ to 4 in our model.

The parameter ${\tt likOfBeingInfected}$ is a multiplier that captures how much more likely it is for an infected individual to be identified as a close contact as compared to a non-infected individual and therefore can be seen as a measure of the effectiveness of contact tracing.
Specifically,
\[
{\tt likOfBeingInfected} = \frac{P(\mbox{close-contact} \ts | \ts  \mbox{infected})}{P(\mbox{close-contact} \ts | \ts  \mbox{non-infected})}.
\]
It would be reasonable to consider settings where ${\tt likOfBeingInfected} \geq 1$ only since if contact tracing is of any value at all it should make randomly chosen infected people more likely to be identified as close-contact than non-infected people. In order to get at least some rough approximation for ${\tt likOfBeingInfected}$, we use the following approach:

We can show that 
\[
{\tt likOfBeingInfected} = \frac{P(\mbox{infected} \ts | \ts  \mbox{close-contact})}{P(\mbox{non-infected} \ts | \ts  \mbox{close-contact})} \div \frac{P(\mbox{infected})}{P(\mbox{non-infected})}.
\]
Thus, we need estimates for the proportion of infected versus non-infected individuals within the non-isolating population as well as the proportion of infected versus non-infected individuals among people identified as close contacts. Of the 1,286 close contacts studied in~\cite{Bi2020}, 98 tested positive. Thus, we estimated the proportion of the infected to non-infected within the close contact group to be $P(\mbox{infected} | \mbox{close-contact})/P(\mbox{non-infected} | \mbox{close-contact}) = 98/1188 = 0.082$.

Next, we estimate $P(\mbox{infected})/P(\mbox{non-infected})$, the proportion of infected to non-infected in the general non-isolating population, using findings of studies done on data collected from Wuhan around the same time the study discussed in~\cite{Bi2020} was conducted. According to~\cite{Wu2020}, in Wuhan, the probability of death after developing symptoms was 1.4\%. Using this estimate and the number of deaths in Wuhan, which is reported as 4,512 by the Johns Hopkins Coronavirus Resource Center~\cite{JHCRC2020}, we can estimate the total number of infected individuals in Wuhan over the course of the pandemic (at least as of now) to be 322,286. In order to estimate the total number of infected individuals at some random time between mid-January and mid-February, we divide this by two and use 161,143 as the average number of infected individuals at a given time. This approximation implicitly assumes a linear build-up of new COVID-positive cases, which is not true for infection spread within a population, but nevertheless serves as a reasonable rough estimate. Then, using 11,000,000 for the population of Wuhan, we estimate $P(\mbox{infected})/P(\mbox{non-infected})$ to be $161,143/(11,000,000-161,143) = 0.015$.

Finally, using our estimates for the infected to non-infected proportion within the close-contacts and within the general non-isolating population, we estimated the ratio of these two proportions to be $0.082/0.015 = 5.47$. Clearly, this is a very rough approximation and it would not be reasonable to reach conclusions based on results only on this estimated value. Therefore, in our analysis we consider scenarios that assume different values around 5.47. Specifically, in the main body of the paper, we report results based on three settings with ${\tt likOfBeingInfected}$ set to 2, 5, and 10, respectively corresponding to increasing levels of contact tracing effectiveness.

\section{Model Validation} \label{section:model-validation}

\vspace{10pt}


We manually calibrated two parameters of the model to roughly match its output to the observed trajectory of the pandemic up until April 15. These two parameters are the $R_0$ value and the initial number of infected asymptomatic-nonisolated individuals on March 2. We ended up with an $R_0$ value of 3.04 and an initial infected asymptomatic-nonisolated population of 676. These starting parameters are tuned based on hospitalization, death, and test data up to June 15. We use testing data from the NC Department of Health and Human Services. We assume that the tests are conducted daily on people in the order of symptomatic, isolated, and finally non-isolated people. 

\begin{figure}[!tbp]
  \centering
  \subfloat
  [Population overview]
  {\includegraphics[scale=0.4]{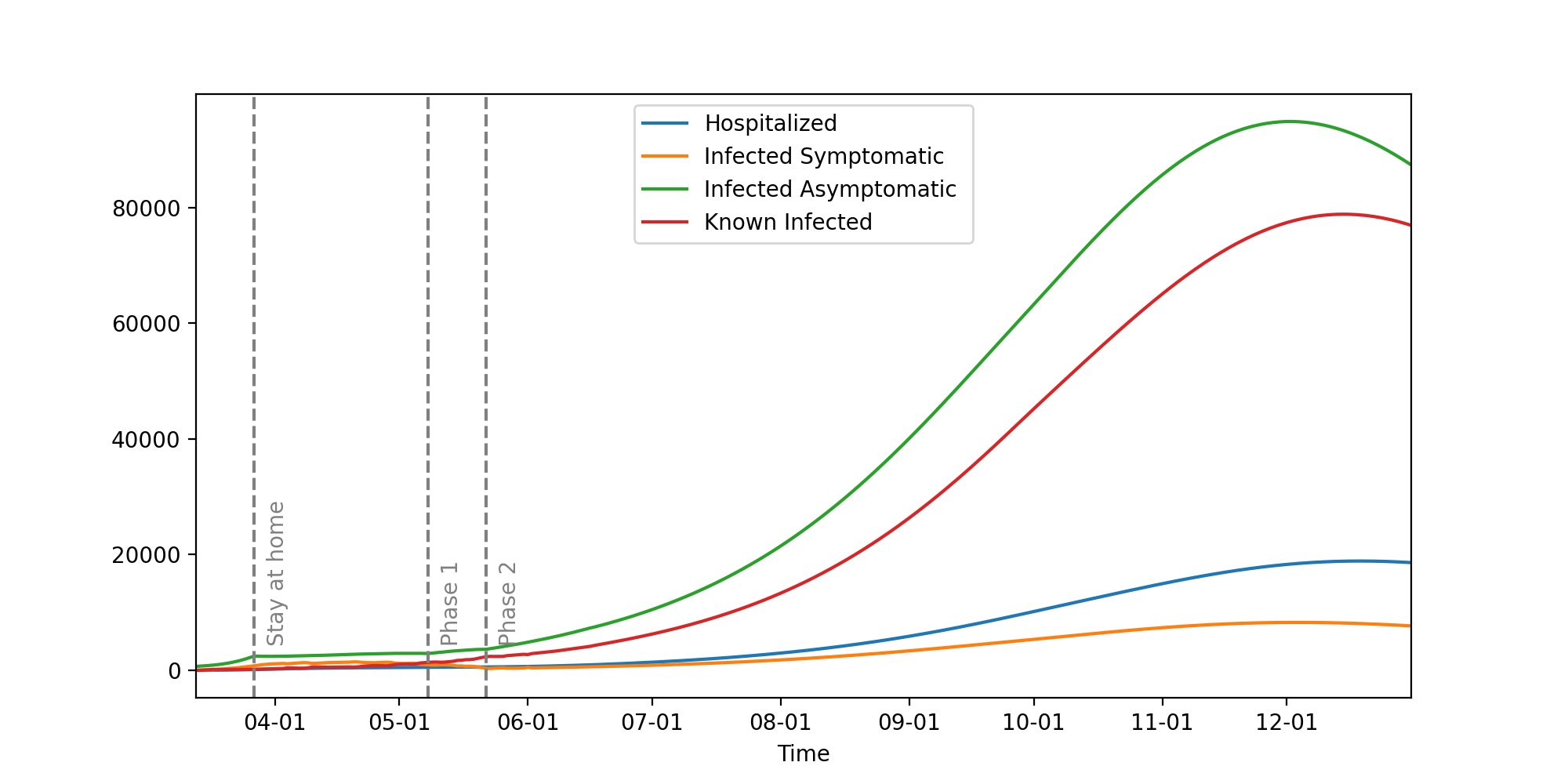}
  \label{fig:overview}}
  \hfill
  \subfloat
  [Cumulative deaths]
  {\includegraphics[scale=0.4]{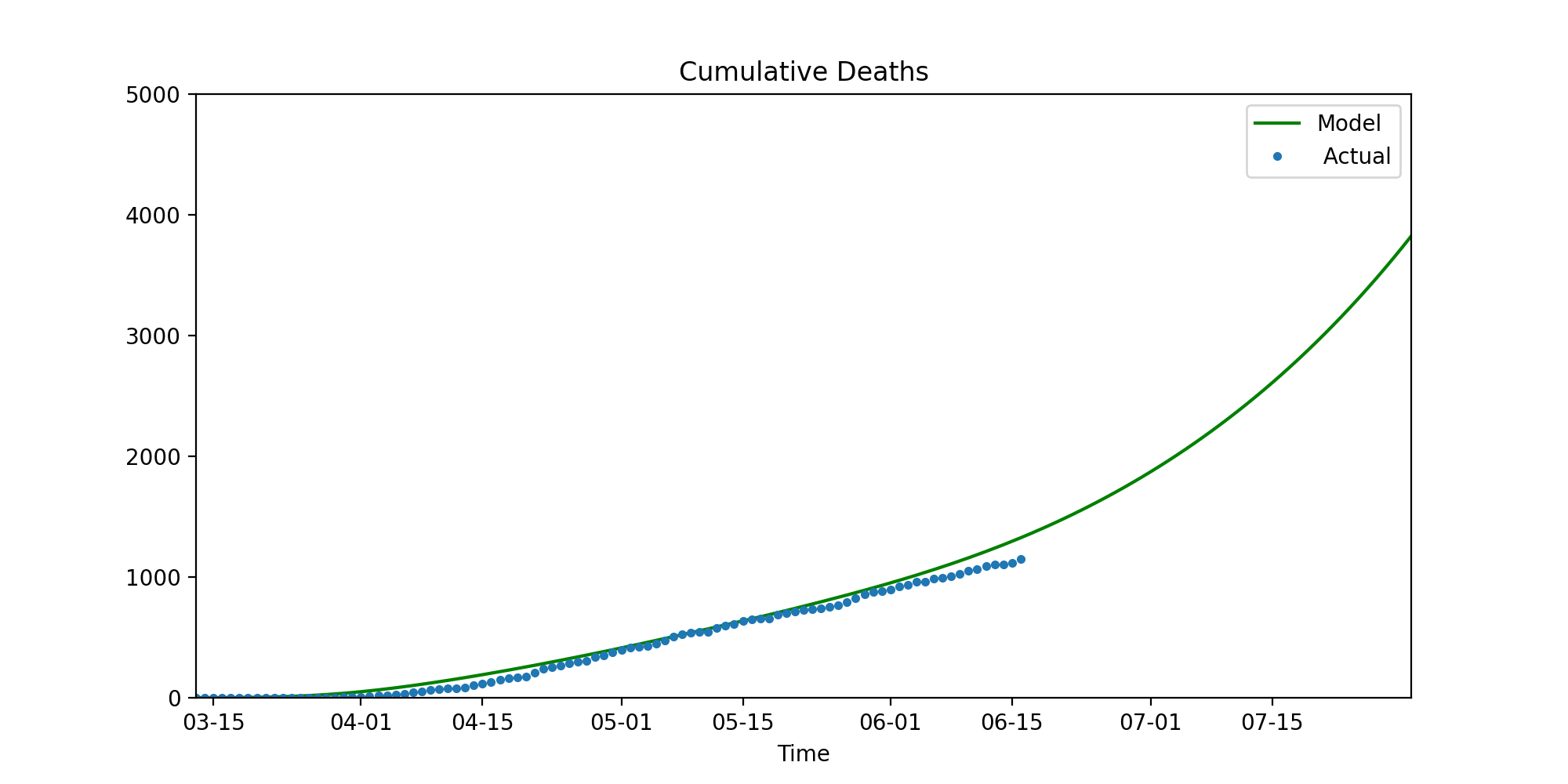}
  \label{fig:cum-deaths}}
  \hfill
  \subfloat
  [Daily new cases]
  {\includegraphics[scale=0.4]{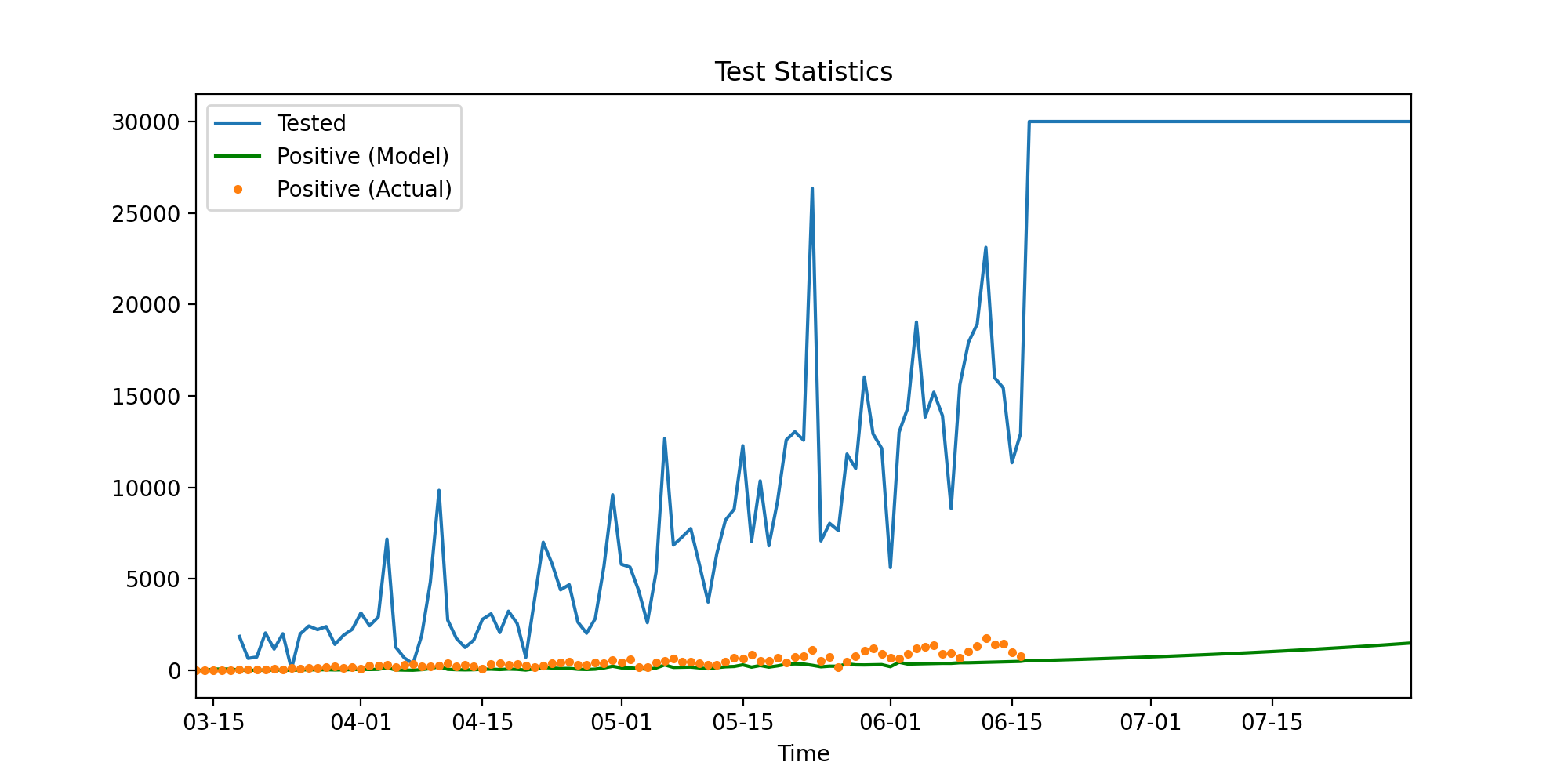}
  \label{fig:cases}}
  \hfill
  \subfloat
  [Daily active hospitalizations]
  {\includegraphics[scale=0.4]{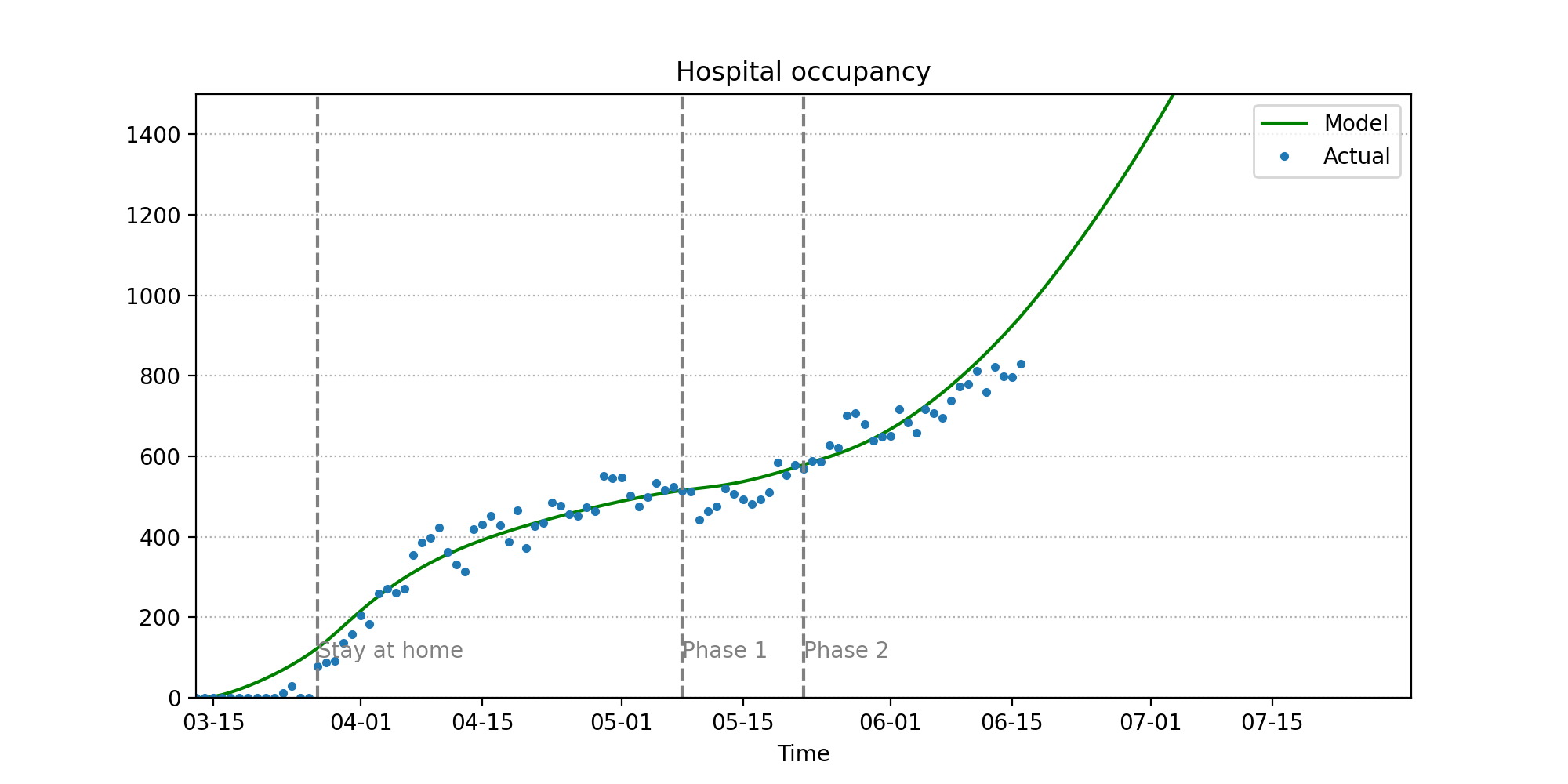}
  \label{fig:new-hosp}}
\caption{Comparison of the actual trajectory of the pandemic and the model.} \label{fig:trajectory}
\end{figure}

The plots in Figure \ref{fig:trajectory} show the epidemic trajectory as predicted by our model. In particular, Figure \ref{fig:overview}  shows the total number of currently hospitalized, symptomatic infected, asymptomatic infected and known infected people on each day as projected by the model. 
In Figure \ref{fig:cum-deaths}, we compare the cumulative number of deaths  predicted by our model with the actual cumulative number of deaths.
In Figure \ref{fig:cases}, we plot the actual daily number of tests conducted and positive cases found in North Carolina. We use the same historical testing capacity in our model. We set the testing capacity to 30,000 for future projections in this plot. Note that the daily number of positive tests in our model fluctuates because we use the actual number of tests performed in
North Carolina until June 15 as the number of tests performed in our model.  However, the number of positive cases identified in our model emerges from the internal behavior of our model. The fit between the model and the actual trajectory in terms of positive cases is reasonably good.
Figure \ref{fig:new-hosp} shows the daily number of active hospitalizations reported by North Carolina against the number of hospitalizations projected by our model. 

\end{document}